\documentclass[10pt]{article}
\usepackage{abstract}
\usepackage{authblk}
\usepackage{samplesty} 
\usepackage{subfigure}
\usepackage{amsmath} 
\usepackage{amsthm}
\usepackage{graphicx}
\usepackage{indentfirst}
\usepackage{color}
\usepackage{cite}
\usepackage{algorithm}
\usepackage{algorithmic}
\usepackage{changepage}
\usepackage{threeparttable}
\newtheorem{rem}{Remark}

\newcommand{\norm}[1]{\left\Vert#1\right\Vert}

\newcommand{\pref}[1]{(\ref{#1})}

\allowdisplaybreaks

\begin{document}

\title{Gregory Solid Construction for Polyhedral Volume Parameterization by Sparse Optimization}

\author[a]{Chuanfeng Hu}
\author[a,b]{Hongwei Lin\thanks{Corresponding author: School of Mathematics, State Key Lab. of CAD\&CG, Zhejiang University, Hangzhou, 310027, China (hwlin@zju.edu.cn)}}
\affil[a]{School of Mathematics, Zhejiang University, Hangzhou, 310027, China}
\affil[b]{State Key Lab. of CAD\&CG, Zhejiang University, Hangzhou, 310027, China}

\renewcommand*{\Affilfont}{\small\it} 
\renewcommand\Authands{, and } 
\date{}

\maketitle

 In isogeometric analysis,
    it is frequently required to handle the geometric models enclosed by four-sided or non-four-sided boundary patches, such as trimmed surfaces.
 In this paper, we develop a Gregory solid based method to parameterize
    those models.
 First, we extend the Gregory patch representation to the trivariate
    Gregory solid representation.
 Second, the trivariate Gregory solid representation is employed to
     interpolate the boundary patches of a geometric model,
     thus generating the polyhedral volume parametrization.
 To improve the regularity of the polyhedral volume parametrization,
    we formulate the construction of the trivariate Gregory solid as a sparse optimization problem,
    where the optimization objective function is a linear combination of some terms,
    including a sparse term aiming to reduce the negative Jacobian area of the Gregory solid.
 Then, the alternating direction method of multipliers (ADMM) is used to
    solve the sparse optimization problem.
 Lots of experimental examples illustrated in this paper demonstrate the
    effectiveness and efficiency of the developed method.

\vspace*{5mm}


\section{Introduction}
\label{sec:introduction}

Isogeometric analysis~\cite{Hughes2005Isogeometric} is an important numerical analysis
     technique that offers the possibility of integrating computer aided design (CAD) and finite element analysis.
While isogeometric analysis requires volumetric representations in some
    cases,
    CAD models are usually defined by boundary representations.
Therefore, to handle the CAD models defined by boundary representations,
    they should be transformed into trivariate volumetric representations.
However, the transformation of boundary representations into volumetric
    representations is not trivial,
    especially when the boundary patches are non-four-sided,
    (e.g. trimmed surfaces),
    and the boundary representation model is homeomorphic to a polyhedron,
    other than hexahedron.

In this paper, we develop a Gregory solid based method to construct the
    polyhedral volume parametrization of CAD models enclosed by boundary patches,
    four-sided or non-four-sided.
Firstly, the polyhedral parametric domain of the CAD model is constructed,
    and then split into several hexahedral sub-domains.
Secondly, a trivariate Gregory solid mapping from the polyhedral parametric
    domain to the CAD model is developed to interpolate the
    boundary patches of the CAD model,
    thus producing the polyhedral volume parametrization of the CAD model.
It is well known that, the volume parametrization that is valid for the
    isogeometric analysis cannot contain self-intersections or folds, i.e.,
    the mapping should be regular.
If the Jacobian~\cite{Knupp2015Achieving} of the mapping does not change
    sign,
    it is regular.
In this paper, the regularity of the Gregory solid mapping is improved by
    solving a sparse optimization problem which minimizes the negative Jocabian area of the Gregory solid.
Finally, the alternating direction method of multipliers
    \cite{Boyd2010Distributed} (ADMM)
    method is employed to solve the sparse optimization problem.

This paper is organized as follows:
In Section~\ref{subsection:related_work},
    we review the related work on Gregory patches,
    generalized barycentric coordinates, and volumetric parametrization.
Section~\ref{sec:gregory_solid} presents the Gregory solid representations.
Section~\ref{sec:optimization} develops the optimization problem for
    improving the parametrization quality.
After some experimental results are demonstrated in
    Section~\ref{sec:results},
    Section~\ref{sec:conclusion} concludes this paper.


\subsection{Related Work}
\label{subsection:related_work}

 Triangular mesh parametrization is a commonly employed technique in curve
    and surface fitting~\cite{Floater1997Parametrization}, texture mapping~\cite{Sander2001Texture}, remeshing~\cite{Alliez1970Recent}, and so on.
 A triangular mesh parametrization constructs a bijective mapping from the
    mesh in three dimension to a planar domain.
 According to the requirements of applications,
    the frequently used mapping methods in mesh parametrization includes
    discrete harmonic mapping~\cite{Floater1997Parametrization}, discrete equiareal mappings~\cite{Hormann2000Mips}, and discrete conformal mapping~\cite{Heras2003An}.
 For more details on triangular mesh parametrization methods and their applications, please refer to~\cite{Floater2005Surface,Sheffer2006Mesh}.

 In the field of trivariate solid modeling,
    the discrete volume parametrization is usually determined by solving a partial differential equation~\cite{Martin2008Volumetric,Martin2010Volumetric} using the finite element method.
 Lin et al.~\cite{Lin2015Constructing} developed the explicit parametric equations that
    maps the vertices of a tetrahedral mesh into a parameter domain,
    thus making the discrete volume parametrization as intuitive and easy to implement as the triangular mesh parametrization methods.
 In the isogeometric analysis,
    the three-dimensional physical domains are usually modeled by trivariate B-spline solids, T-spline solids, and subdivision solids, etc.,
    which are generally constructed by filling the CAD models with boundary representation.
 Wang et al.~\cite{Wang2013Trivariate} proposed a method that constructs a T-spline
    solid from boundary triangulations with arbitrary genus topology by the polycube mapping.
 In 2013, Xu et al.~\cite{Xu2013Analysis} presented a method to obtain
    analysis-suitable trivariate NURBS and improve the mesh quality.
 Zhang et al.~\cite{Zhang2013Conformal} developed an approach for volumetric
    T-spline construction that considers boundary layers.
 In 2014, an optimization-based approach was developed to
     generate the B-spline solid with positive Jacobian values from boundary-represented model with six boundary surfaces~\cite{Wang2014An}.
 In 2015, Lin et al.~\cite{Lin2015Constructing} presented a discrete volume
    parametrization method for tetrahedral mesh models with six boundary surfaces,
    and an iterative fitting algorithm for constructing a B-spline solid.
 In 2018, Lin et al.~\cite{Lin2018Trivariate} proposed a method to construct
    a trivariate B-spline solid by pillow operation and geometric iterative fitting.

 To parameterize the mesh vertices of a mesh model with complicated shape
    for mesh deformation,
    the generalized barycentric coordinates were developed.
 In 2002, Meyer et al.~\cite{Meyer2002Generalized} presented an easy computation method of a
     generalized form of barycentric coordinates for irregular, convex n-sided polygons, not only for triangles.
 Moreover, the mean-value coordinates~\cite{Floater2003Mean} was developed for both
    convex and concave polygons,
    and generalized to 3D polyhedral domains~\cite{Floater2005Mean}.
 In 2007, Joshi et al.~\cite{Joshi2007Harmonic} introduced the harmonic coordinates based
    on the solutions of the Laplace's equation,
    which can work on convex and concave polyhedrons.
 Because this method does not have a closed form solution,
    the boundary conditions and the solutions must be defined for every particular case.
 In 2008, Lipman et al.~\cite{Lipman2008Green} presented the Green coordinates based on
    the solution of the Green's function,
    which can produce a conformal mapping in 2D and a quasi-conformal mapping in 3D.
 Note that, nearly all of the generalized barycentric coordinates methods
    require that the input models are solid.
 However, the input models handled in this paper are hollowed,
    enclosed by boundary patches.
 Therefore, the generalized barycentric coordinates methods cannot be
    directly used to parameterize the hollowed models handled in this paper.

 In this paper, we developed the representation of the trivariate Gregory
    solid,
    and employed it to fill the models enclosed by boundary patches,
    thus generating the polyhedral volume parametrization of the input models.
 The Gregory patch~\cite{Chiyokura1983Design,Chiyokura1986Localized} arose from the Gregory's method~\cite{Gregory1974SMOOTH},
 which produces the 8 inner control points from four boundary edges and four corner points, one pair per corner.
 And then, the four pairs of inner control points are blended so that the
    generated patch interpolates the boundary straight line segments.
 Similarly, a triangular Gregory patch can be constructed
    using the method proposed in~\cite{Longhi1987Interpolating}.
 Moreover, Wang et al.~\cite{Wang2004Non} defined the Gregory patch as a
    mapping from an $n$-sided
    parametric domain with straight line boundaries to an $n$-sided parametric domain of a trimmed surface,
    and non-self-overlapping structured grids can be generated on it,
    as well as the trimmed surface.


\section{Gregory Solid Representation}
\label{sec:gregory_solid}

 Suppose we are given a \emph{physical domain}, i.e., a curved and hollowed
    polyhedron $\mathcal{H}$ enclosed by boundary patches.
 The boundary patches can be any type of parametric surfaces,
    e.g., parameterized triangular meshes, trimmed surfaces, and so on.
 In this paper, we develop a Gregory solid representation,
    and use it to fill the hollowed polyhedron $\mathcal{H}$,
    thus generating the polyhedral volume parametrization of the polyhedral physical domain.
 It should be pointed out that,
    the construction of the Gregory solid requires that each corner of the given polyhedron is adjacent to just \emph{three} boundary patches.
 So, in the following,
    the given polyhedral physical domain $\mathcal{H}$ is supposed to satisfy the requirement.


\begin{figure}[!htb]
\begin{center}
\includegraphics[width=0.4\columnwidth]{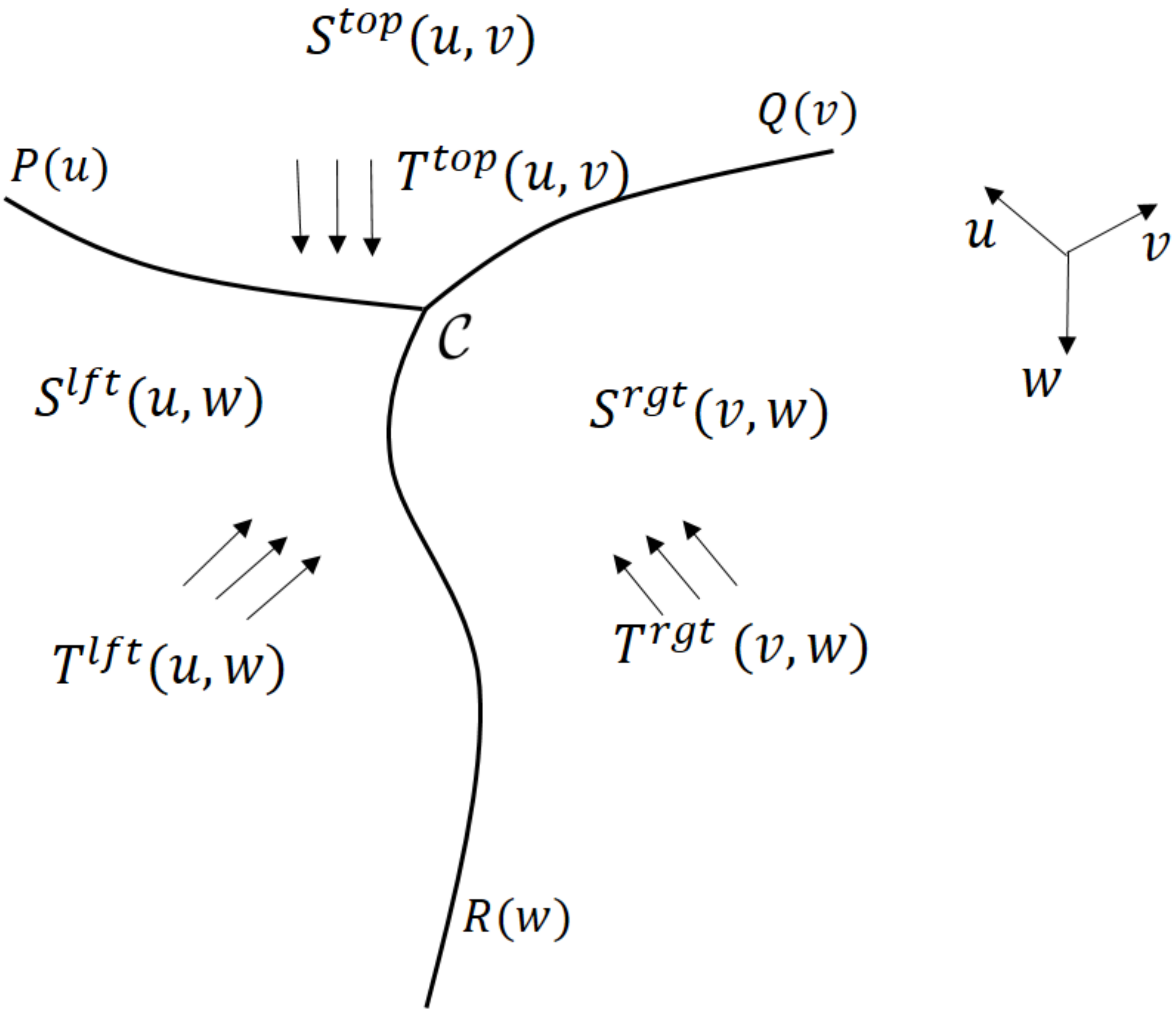}\\
\caption{\small The definition of a Gregory corner interpolator.}
\label{fig:corner_interpolator}
\end{center}
\end{figure}

 \subsection{Gregory corner interpolator}
 \label{subsection:corner_interpolator}

 The Gregory corner interpolator is defined at each corner of the given
    polyhedral physical domain $\mathcal{H}$.
 Suppose $P(u),\ Q(v),\ \text{and}\ R(w), (0 \le u \le 1,\ 0 \le v \le 1,\ 0 \le w \le 1)$ are three boundary curves adjacent
    to a corner $\mathcal{C}$ of $\mathcal{H}$ (refer to Fig.~\ref{fig:corner_interpolator}).
 Whatever the type of the boundary patches adjacent to the corner
    $\mathcal{C}$ is,
    we now rewrite them in the form of parametric patches, i.e.,
    $$
    S^{top}(u,v),\ S^{lft}(u,w),\ \text{and}\ S^{rgt}(v,w),\
    (0 \le u \le 1,\ 0 \le v \le 1,\ 0 \le w \le 1)
    $$
    which interpolate the three boundary curves (Fig. 1),
    $$
    S^{lft}(u,0) = S^{top}(u,0) = P(u),\
    S^{top}(0,v) = S^{rgt}(v,0) = Q(v),\
    S^{rgt}(0,w) = S^{lft}(0,w) = R(w).
    $$
 Moreover, we assign the cubic B-spline vector functions
    $\bar{P}^{lft}(u),\ \bar{P}^{rgt}(u)$ on the boundary curve $P(u)$,
    the vector functions $\bar{Q}^{lft}(v),\ \bar{Q}^{rgt}(v)$ on the boundary curve $Q(v)$,
     and the vector functions $\bar{R}^{lft}(w),\ \bar{R}^{rgt}(w)$ on the boundary curve $R(w)$, respectively.
 The vector functions should satisfy the \emph{compatibility conditions},
 \begin{equation*}
   \bar{R}^{rgt}(0) = \bar{Q}^{lft}(0),\ \bar{P}^{rgt}(0) = \bar{R}^{lft}(0),\
   \bar{Q}^{rgt}(0) = \bar{P}^{lft}(0).
 \end{equation*}
 In general, they are generated by approximating the tangent vector functions
    of the boundary patches at the boundary curves, i.e.,
 \begin{itemize}
   \item $\bar{P}^{lft}(u)$ approximates $\frac{d S^{lft}(u,w)}{dw}\bigg|_{w=0}$,
    and $\bar{P}^{rgt}(u)$ approximates $\frac{d S^{top}(u,v)}{dv}\bigg|_{v=0}$;
   \item $\bar{Q}^{lft}(v)$ approximates $\frac{d S^{top}(u,v)}{du}\bigg|_{u=0}$,
    and $\bar{Q}^{rgt}(v)$ approximates $\frac{d S^{rgt}(v,w)}{dw}\bigg|_{w=0}$;
   \item $\bar{R}^{lft}(w)$ approximates $\frac{d S^{rgt}(v,w)}{dv}\bigg|_{v=0}$,
    and $\bar{R}^{rgt}(w)$ approximates $\frac{d S^{lft}(u,w)}{du}\bigg|_{u=0}$.
 \end{itemize}
 Then, we construct three bi-cubic B-spline vector functions
    $T^{top}(u,v),\ T^{lft}(u,w),\ \text{and}\ T^{rgt}(v,w),$
    that interpolate the vector functions defined above.
 Specifically,
 \begin{itemize}
   \item $T^{lft}(u,w)$ interpolates $\bar{R}^{lft}(w)$ and $\bar{P}^{rgt}(u)$;
   \item $T^{top}(u,v)$ interpolates $\bar{P}^{lft}(u)$ and $\bar{Q}^{rgt}(v)$;
   \item $T^{rgt}(v,w)$ interpolates $\bar{Q}^{lft}(v)$ and $\bar{R}^{rgt}(w)$.
 \end{itemize}
 The bi-cubic B-spline vector functions
    $T^{top}(u,v),\ T^{lft}(u,w),\ \text{and}\ T^{rgt}(v,w)$
    can be written as,
 \begin{equation}
 \label{eq:vector_functions}
 \begin{split}
    T^{top}(u,v) & =
     \sum\limits_{i=0}^{n_u}\sum\limits_{j=0}^{n_v}N_{i,p}(u)N_{j,q}(v)X^{top}_{ij},\\
    T^{lft}(u,w) & =
     \sum\limits_{i=0}^{n_u}\sum\limits_{k=0}^{n_w}N_{i,p}(u)N_{k,r}(w)X^{lft}_{ik},\\
    T^{rgt}(v,w) & =
     \sum\limits_{j=0}^{n_v}\sum\limits_{k=0}^{n_w}N_{j,q}(v)N_{k,r}(w)X^{rgt}_{jk},
 \end{split}
 \end{equation}
    where $X^{top}_{ij}$, $X^{lft}_{ik}$ and $X^{rgt}_{jk}$, $i=0,1,...,n_u,j=0,1,...,n_v,k=0,1,...,n_w$ are control points,
    and $N_{i,p}(u)$, $N_{j,q}(v)$, $N_{k,r}(w)$ are the basis of B-splines of degree $p$ in the $u$, degree $q$ in the $v$ and degree $r$ in the $w$.
    In these control points, only the control points of the vector functions
    $$\bar{P}^{lft}(u),\ \bar{P}^{rgt}(u),\ \bar{Q}^{lft}(v),\ \bar{Q}^{rgt}(v),\ \bar{R}^{lft}(w),\ \text{and}\ \bar{R}^{rgt}(w)$$ are known,
    and the other control points are unknown.
 They will be taken as variables in the optimization procedure stated in
    Section~\ref{sec:optimization},
    and determined by solving the optimization problem.

 \begin{rem}[Construction of the initial patches]
 \label{rmk:initial_patch}
 In solving the optimization problem developed in Section
    \ref{sec:optimization},
    the initial patches of $T^{top}(u,v),\ T^{lft}(u,w)$, and $T^{rgt}(v,w)$ are required.
 Take the construction of the initial representation of $T^{top}(u,v)$ as an
    example.
 As stated above, $T^{top}(u,v)$ should interpolate the two curves
    $\bar{P}^{lft}(u)$ and $\bar{Q}^{rgt}(v)$,
    which are two boundary curves of $T^{top}(u,v)$.
 In order to produce the other two boundary curves of $T^{top}(u,v)$,
    we first construct a corner,
    \begin{equation*}
        \mathcal{C}_p = \frac{\bar{P}^{lft}(1) + \bar{Q}^{rgt}(1)}{2} +
            2 \left(
                    \frac{\bar{P}^{lft}(1) + \bar{Q}^{rgt}(1)}{2}
                    -
                    \bar{P}^{lft}(0)
              \right),
    \end{equation*}
    and then, connect the two corners $\bar{P}^{lft}(1)$ and $\mathcal{C}_p$,
    and the two corners $\bar{Q}^{rgt}(1)$ and $\mathcal{C}_p$,
    thus generating two line segments as the other two boundary curves of $T^{top}(u,v)$.
 In this way, we get four boundary curves,
    and a bilinear patch can be generated by bilinear interpolation to the four boundary curves.
 Moreover, by degree elevation, the bilinear patch becomes a bi-cubic
    B-spline patch,
    which can be taken as the initial patch of $T^{top}(u,v)$.
 The initial patches of $T^{lft}(u,w)$ and $T^{rgt}(v,w)$ can be constructed
    in the similar manner.
 \end{rem}

 In conclusion, the Gregory corner interpolator with respect to the corner
    $\mathcal{C}$ that interpolates
 $$\{S^{top}(u,v), S^{lft}(u,w), S^{rgt}(v,w), T^{top}(u,v), T^{lft}(u,w), T^{rgt}(v,w)\}$$
    can be represented as,
 \begin{equation}
 \label{eq:corner_interpolator}
 \begin{split}
 R&(u,v,w) = [1 \quad w]{
  \left[
  \begin{array}{c}
   S^{top}(u,v) \\
   T^{top}(u,v) \\
  \end{array}
  \right]}
  + [1 \quad v]{
  \left[
  \begin{array}{c}
  S^{lft}(u,w) \\
  T^{lft}(u,w) \\
  \end{array}
  \right ]}
  + [1 \quad u]{
  \left[
  \begin{array}{c}
  S^{rgt}(v,w) \\
  T^{rgt}(v,w) \\
  \end{array}
  \right ]} \\
  &- [1 \quad u]{
  \left[
  \begin{array}{cc}
    S^{lft}(0,w) & T^{lft}(0,w)\\
    T^{rgt}(0,w) & \frac{vT^{lft}_u(0,w)+uT^{rgt}_v(0,w)}{u+v} \\
  \end{array}
  \right]}{
  \left[
  \begin{array}{c}
  1 \\
  v \\
  \end{array}
  \right]}\\
  &- [1 \quad v]{
  \left[
  \begin{array}{cc}
  S^{top}(u,0) & T^{top}(u,0) \\
  T^{lft}(u,0) & \frac{vT^{lft}_w(u,0)+wT^{top}_v(u,0)}{v+w}\\
  \end{array}
  \right ]}{
  \left[
  \begin{array}{c}
  1 \\
  w \\
  \end{array}
  \right ]}\\
  &- [1 \quad w]{
  \left[
  \begin{array}{cc}
  S^{rgt}(v,0) & T^{rgt}(v,0) \\
  T^{top}(0,v) & \frac{uT^{rgt}_w(v,0)+wT^{top}_u(0,v)}{u+w}\\
  \end{array}
  \right ]}{
  \left[
  \begin{array}{c}
  1 \\
  u \\
  \end{array}
  \right ]}\\
  &+ \mathcal{T}_{ijk}u^i v^j w^k,
 \end{split}
 \end{equation}
 where the Einstein's summation convention is applied in the last term,
 \begin{equation*}
 u^0 = 1,\ u^1 = u,\ v^0 = 1,\ v^1 = v,\ w^0 = 1,\ w^1 = w,
 \end{equation*}
 and $\mathcal{T}_{ijk}$ is a 3-order tensor with elements,
 \begin{adjustwidth}{2cm}{1cm}
 $\mathcal{T}_{000}=S^{top}(0,0)$, \quad
 $\mathcal{T}_{100}=T^{rgt}(0,0)$, \quad
 $\mathcal{T}_{010}=T^{lft}(0,0)$, \quad
 $\mathcal{T}_{001}=T^{top}(0,0)$,\\
 $\mathcal{T}_{110}=\frac{vT^{lft}_u(0,0)+uT^{rgt}_v(0,0)}{u+v}$, \quad
 $\mathcal{T}_{011}=\frac{vT^{lft}_w(0,0)+wT^{top}_v(0,0)}{v+w}$, \quad
 $\mathcal{T}_{101}=\frac{uT^{rgt}_w(0,0)+wT^{top}_u(0,0)}{u+w}$,\\
 $\mathcal{T}_{111}=\frac{ uv(uT^{rgt}_{vw}(0,0) + vT^{lft}_{uw}(0,0)) + uw(uT^{rgt}_{wv}(0,0) + wT^{top}_{uv}(0,0)) + vw(vT^{lft}_{wu}(0,0) + wT^{top}_{vu}(0,0))}{uv(u+v) + uw(u+w) + vw(v+w)}$.\\
\end{adjustwidth}
 Here, $T^{rgt}_v(v,w)$ denotes the first order partial derivative
    $\frac{\partial T^{rgt} (v,w)}{\partial v}$,
    $T^{rgt}_{wv}(v,w)$ denotes the second order partial derivative
    $\frac{\partial^2 T^{rgt}(v,w)}{\partial w \partial v}$,
    and so on.
 It is easy to be validated that,
    the Gregory corner interpolator~\pref{eq:corner_interpolator} interpolates the three boundary patches adjacent to the corner $\mathcal{C}$, i.e.,
    $$
    R(u,v,0)=S^{top}(u,v), R(u,0,w)=S^{lft}(u,w), \text{and}\ R(0,v,w)=S^{rgt}(v,w),
    $$
    and its partial derivatives satisfy,
   $$
    \frac{\partial R(u,v,w)}{\partial w}\bigg|_{w=0}= T^{top}(u,v), \quad \frac{\partial R(u,v,w)}{\partial v}\bigg|_{v=0}= T^{lft}(u,w), \quad \frac{\partial R(u,v,w)}{\partial u}\bigg|_{u=0}= T^{rgt}(v,w).
   $$

 As stated above, the Gregory corner interpolator
    \pref{eq:corner_interpolator} is defined at each corner of the given polyhedral physical domain $\mathcal{H}$.
 The Gregory solid representation is the weighted sum of the Gregory corner
    interpolators at all of the corners of $\mathcal{H}$,
    which will be presented in Section~\ref{subsection:Gregory_solid}.

\subsection{Parametric domain}
\label{subs:parametric_domain}

\begin{figure}[!htb]
\begin{center}
\includegraphics[width=0.5\columnwidth]{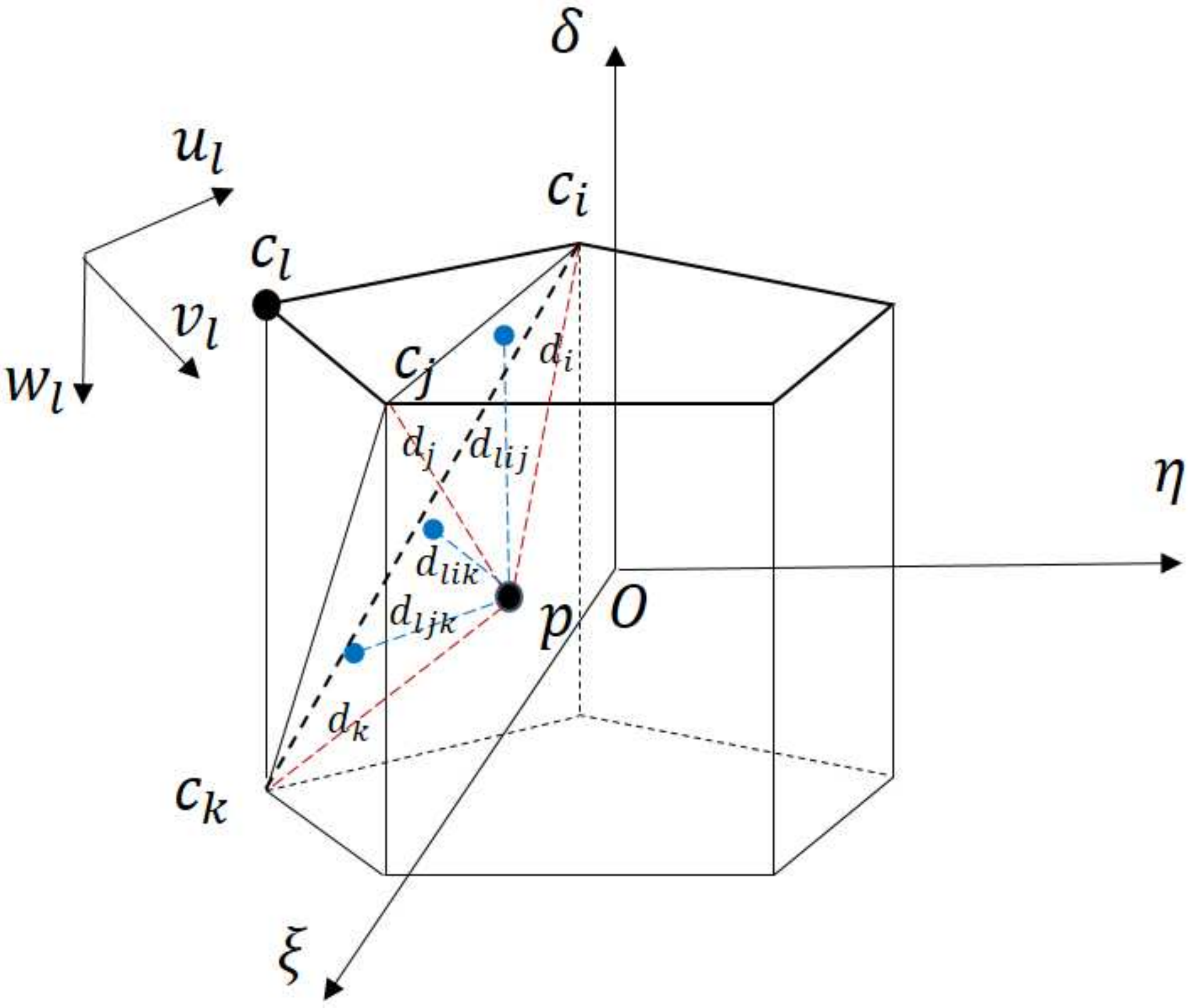}\\
\caption{\small
    The calculation of the parameter values $(u_l, v_l, w_l)$ of the point $p$ with respect to the corner $c_l$ of the pentagonal prism parametric domain $\mathcal{P_{G}}$.
    }
\label{fig:parametric_domain}
\end{center}
\end{figure}

 Before constructing the Gregory solid $\mathcal{G}$ that interpolates the
    boundary patches of the polyhedral physical domain $\mathcal{H}$,
    its parametric domain $\mathcal{P_G}$ should be firstly specified.
 The parametric domain $\mathcal{P_G}$ is a polyhedron in the
    $\xi-\eta-\delta$ parametric space (refer to Fig.~\ref{fig:parametric_domain}),
    which is determined by the number of the boundary patches of the physical domain $\mathcal{H}$.
 For example, if $\mathcal{H}$ has four boundary patches,
    the parametric domain $\mathcal{P_G}$ of the Gregory solid is a tetrahedron.
 In Fig.~\ref{fig:parametric_domain},
    a pentagonal prism parametric domain is illustrated,
    and the corresponding physical domain $\mathcal{H}$ has seven boundary patches.
 In our implementation, the edges of the parametric domain $\mathcal{P_G}$
    (a polyhedron) are unit length.

 Given a point $p=(\xi,\eta,\delta) \in \mathcal{P_G}$,
    it has parameter values in each Gregory corner interpolator~\pref{eq:corner_interpolator}.
 So, the parameter values for the point $p=(\xi,\eta,\delta)$ should be
    defined with respect to each corner of $\mathcal{P_G}$.
 Refer to Fig.~\ref{fig:parametric_domain}, the parameter values
    $(u_l,v_l,w_l)$ of the point $p=(\xi,\eta,\delta)$ with respect to the corner $c_l$ is constructed according to the following manner.
 Consider the tetrahedron $c_l-c_i c_j c_k$.
 On one hand, denote the distances from the point $p=(\xi,\eta,\delta)$ to
    the planes determined by the triangular faces $\triangle c_l c_i c_k$, $\triangle c_l c_i c_j$, and $\triangle c_l c_j c_k$ are $d_{lik},\ d_{lij},\ d_{ljk}$, respectively.
 On the other hand, denote the distances from the point $p=(\xi,\eta,\delta)$
    to the three corners $c_i,\ c_j$, and $c_k$ are $d_i,\ d_j$, and $d_k$, respectively.
 The parameter values $(u_l,v_l,w_l)$ of the point $p=(\xi,\eta,\delta)$ with
    respect to the corner $c_l$ is defined as,
    $$
    (u_l,v_l,w_l) =
    \left(
    \frac{d_{ljk}}{d_{ljk}+d_i},
     \frac{d_{lik}}{d_{lik}+d_j},
     \frac{d_{lij}}{d_{lij}+d_k}
    \right).
    $$
 It can be easily checked that,
 \begin{itemize}
   \item if the point $p$ is at the corner $c_l$,
            we have $(u_l,v_l,w_l)=(0,0,0)$;
   \item if the point $p$ is at the corner $c_i$,
            we have $(u_l,v_l,w_l)=(0,0,1)$;
   \item if the point $p$ is at the corner $c_j$,
            we have $(u_l,v_l,w_l) = (1,0,0)$;
   \item if the point $p$ is at the corner $c_k$,
            we have $(u_l,v_l,w_l) = (0,1,0)$;
   \item if the point $p$ is in the line $c_l c_i$,
            we have $u_l = v_l = 0$;
   \item if the point $p$ is in the line $c_l c_j$,
            we have $v_l = w_l = 0$;
   \item if the point $p$ is in the line $c_l c_k$,
            we have $u_l = w_l = 0$;
   \item if the point $p$ is on the plane determined by
            $\triangle c_l c_i c_k$,
            we have $u_l = 0$;
   \item if the point $p$ is on the plane determined by
            $\triangle c_l c_i c_j$,
            we have $v_l = 0$;
   \item if the point $p$ is on the plane determined by
            $\triangle c_l c_j c_k$,
            we have $w_l = 0$.
 \end{itemize}

 \subsection{Gregory solid representation}
 \label{subsection:Gregory_solid}

 Given a polyhedral physical domain $\mathcal{H}$ with $n$ corners.
 In this section,
    we will develop the representation of the Gregory solid $\mathcal{G}$
    that fills the physical domain $\mathcal{H}$,
    and interpolates its boundary patches at the same time.
 Accordingly, the parametric domain $\mathcal{P_G}$ of $\mathcal{G}$ also
    has $n$ corners $c_l, l =0,1,\cdots,n-1$.
 Moreover, suppose it has $m$ faces $f_i, i=0,1,\cdots,m$.
 For a point $p=(\xi,\eta,\delta) \in \mathcal{P_G}$,
    denoting $d(p,f_i)$ as the distance from the point $p$ to the face $f_i, i=0,1,\cdots,m$,
    the weight function $W_l(p)$ for the corner $c_l, l=0,1,\cdots,n-1$ can be defined as,
    \begin{equation}
    \label{eq:weight}
        W_l(p) = \frac{\prod_{\text{$f_i$ is not adjacent to $c_l$}}
                                d^2(p,f_i)}
                      {\sum_{j=0}^{n-1}
                       \prod_{\text{$f_i$ is not adjacent to $c_j$}}
                                d^2(p,f_i)},\
                      p \in \mathcal{P_G}.
    \end{equation}
 Then, the Gregory solid
    $\mathcal{G}(p): \mathcal{P_G} \rightarrow \mathcal{H}$
    can be defined as the weighted sum of the $n$ corner interpolator functions,
 \begin{equation}
 \label{eq:gregory_solid}
    \mathcal{G}(p) = \sum_{l=0}^{n-1}
                    W_l(p) R_l(u_l(p), v_l(p), w_l(p)),
 \end{equation}
 where $R_l(u_l(p), v_l(p), w_l(p))$ is the Gregory corner interpolator
    \pref{eq:corner_interpolator} to the $l^{th}$ corner
    $c_l, l=0,1,\cdots, n-1$.

 It should be pointed out that,
    $W_l(p) = 1$ if $p$ is at the corner $c_l$,
    and $W_l(p)$ is zero if $p$ is on the faces not adjacent to the corner $c_l$.
 Therefore, the Gregory solid $\mathcal{G}(p)$ interpolates the boundary
    patches of the physical domain $\mathcal{H}$.
 Specifically, if $p_1$ is on a face of the parametric domain
    $\mathcal{P_G}$,
    then there is a point $q_1$ on a corresponding boundary patch of $\mathcal{H}$ such that $\mathcal{G}(p_1) = q_1$.
 On the contrary, if there is a point $q_2$ on a boundary patch of
    $\mathcal{H}$,
    then there exists a point $p_2$ on a corresponding face of $\mathcal{P_G}$ such that $\mathcal{G}(p_2) = q_2$.

\begin{figure}[!htb]
  \begin{center}
  \subfigure[]{
    \label{subfig:hex}
    \includegraphics[width=0.25\textwidth]{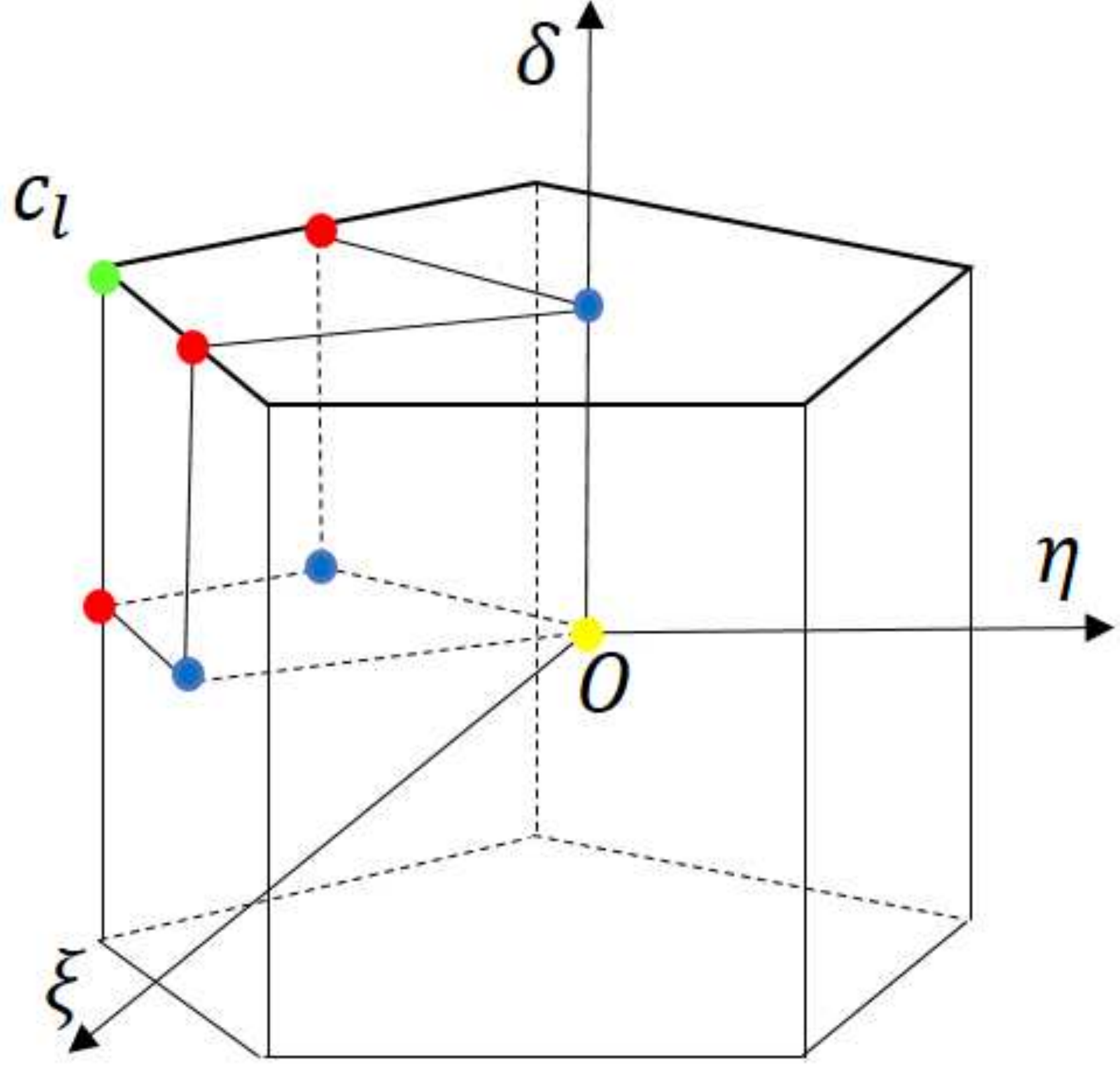}}
  \subfigure[]{
    \label{subfig:polyhedron}
    \includegraphics[width=0.20\textwidth]{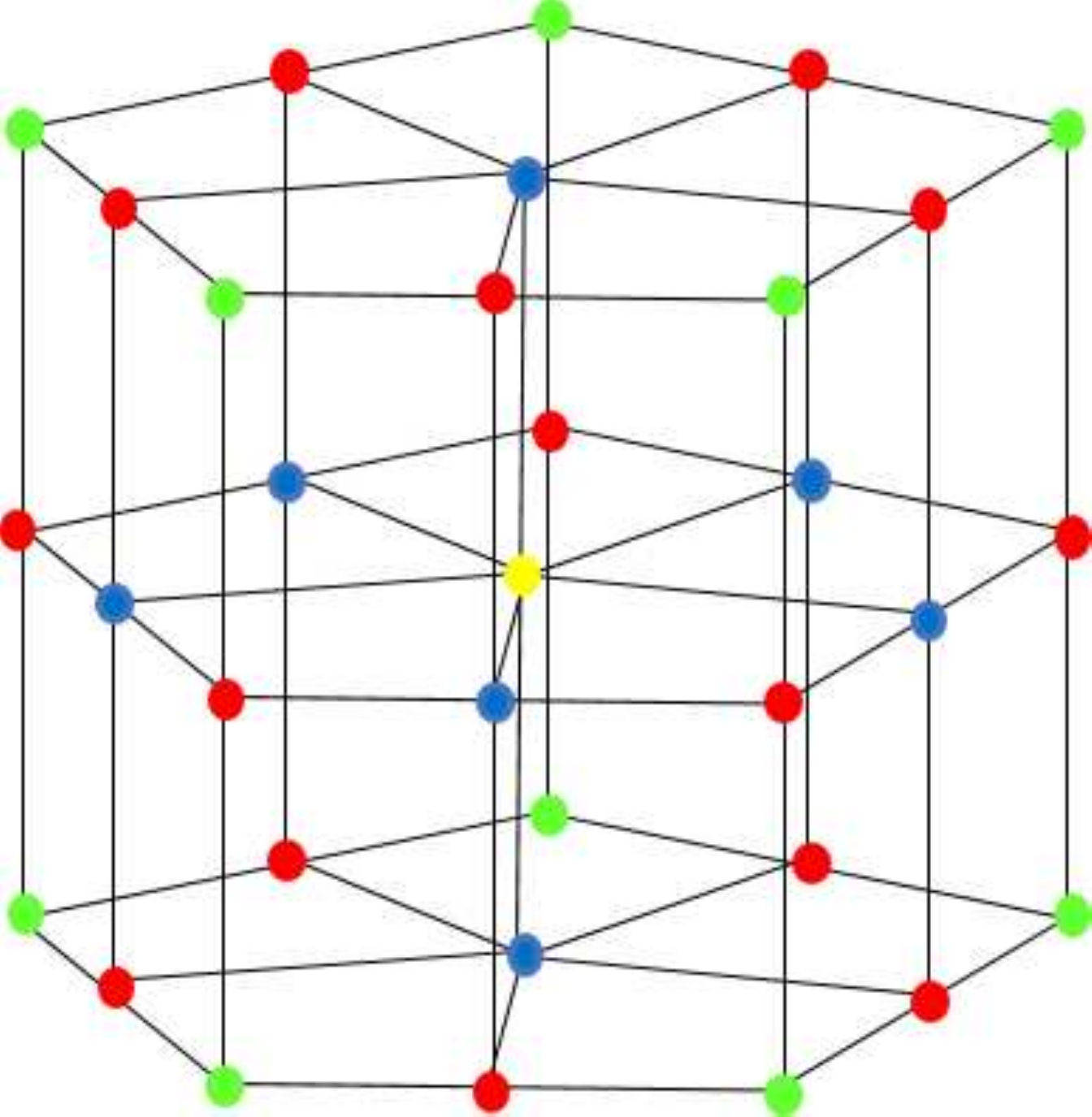}}
  \subfigure[]{
    \label{subfig:grid}
    \includegraphics[width=0.24\textwidth]{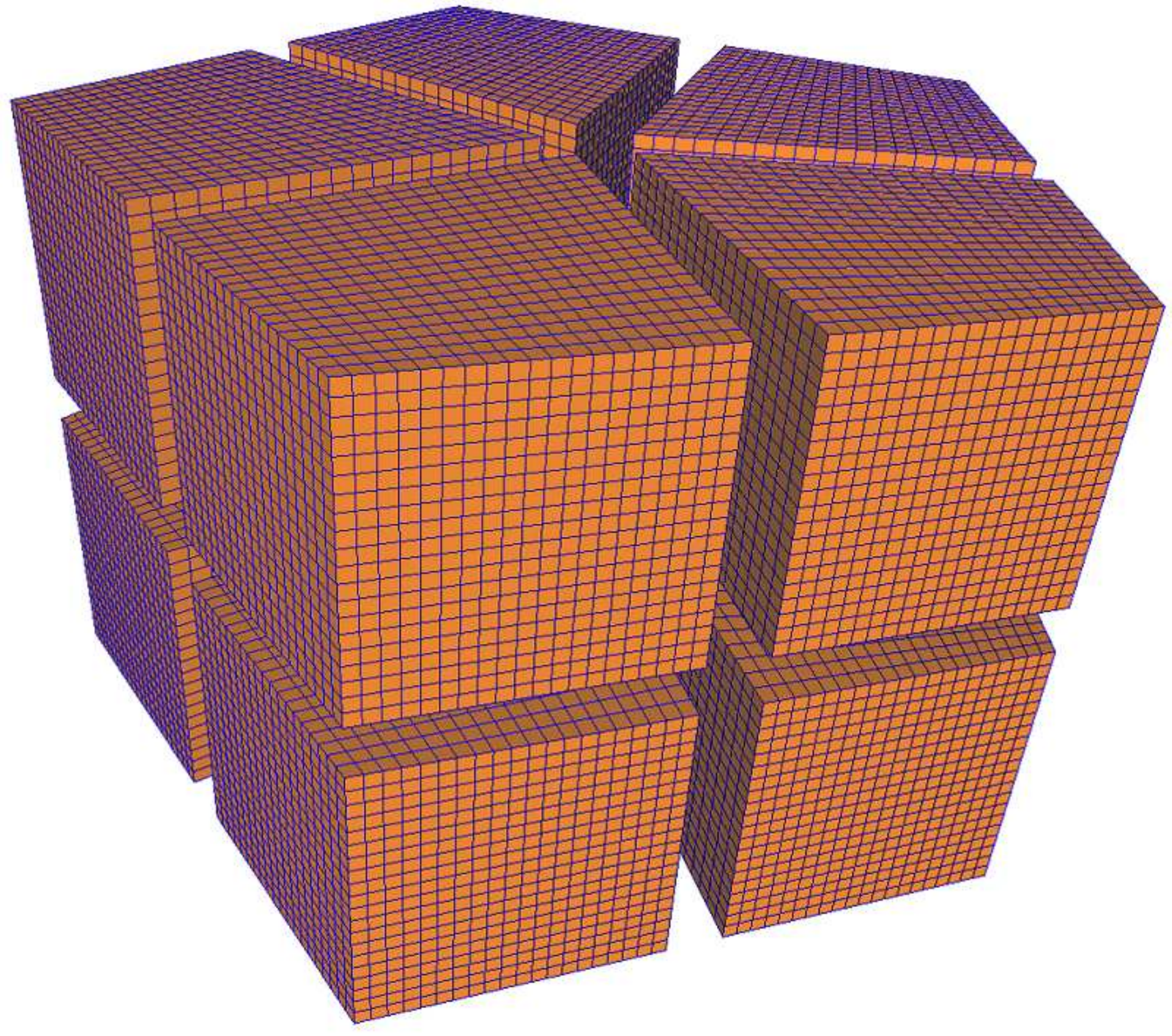}}
  \subfigure[]{
    \label{subfig:seperatedhead}
    \includegraphics[width=0.12\textwidth]{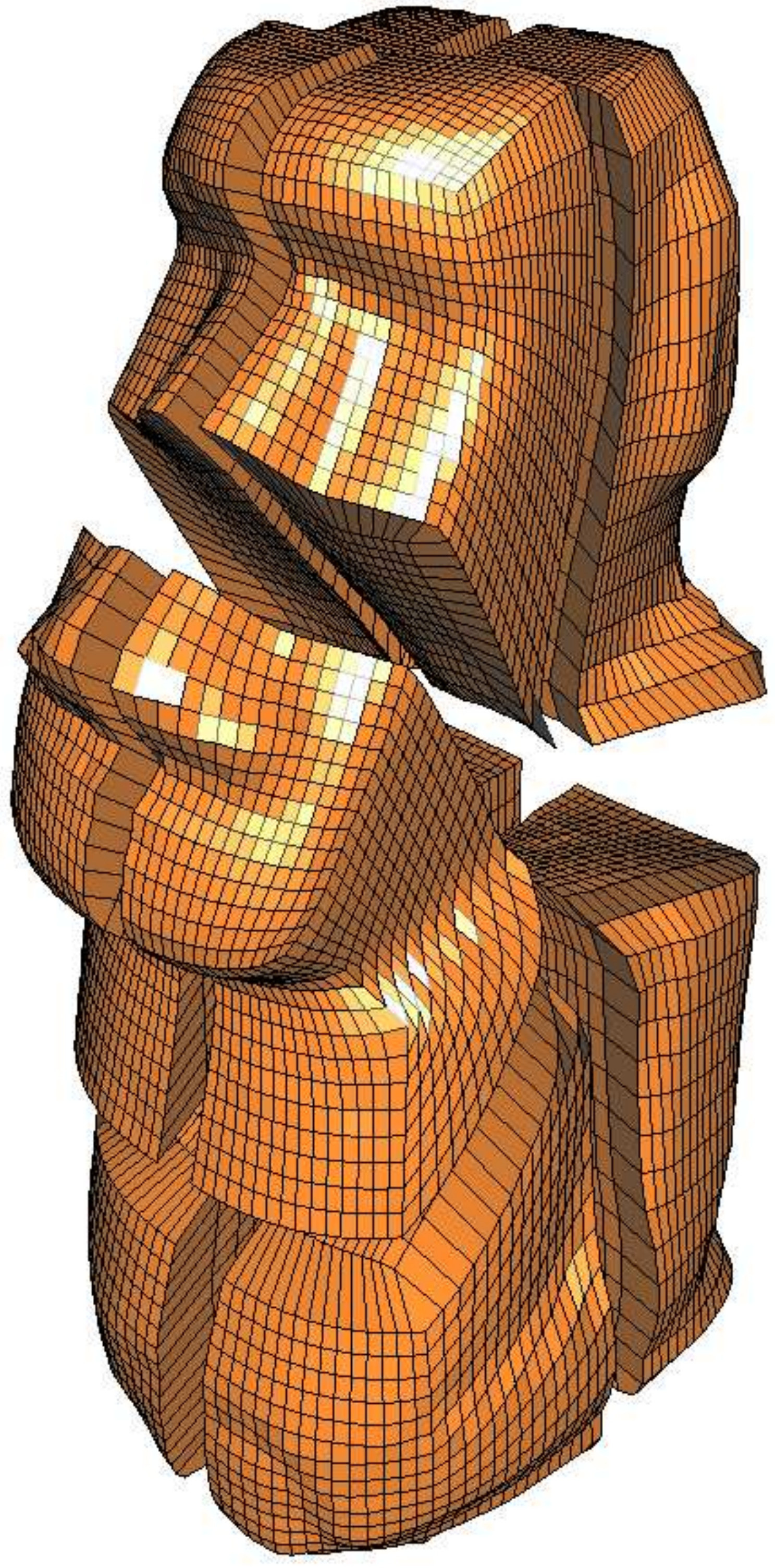}}
      \caption{\small Parametric grid generation.
      (a) Dual operation for a corner.
      (b) Parametric domain after dual operation.
      (c) Parametric domain discretization (separated).
      (d) Parametric grid in the physical domain (separated).}
   \label{fig:multi grid}
  \end{center}
\end{figure}

 \subsection{Parametric grid generation}
 \label{subs:parametric_grid}

 Now, we have constructed a Gregory solid
    $\mathcal{G}(p),\ p \in \mathcal{P_G}$ which fills the
    polyhedral physical domain $\mathcal{H}$,
    and interpolates its boundary patches.
 The Gregory solid $\mathcal{G}(p)$ can be considered as a mapping from its
    parametric domain $\mathcal{P_G}$ to the physical domain $\mathcal{H}$,
    i.e., $\mathcal{G}(p) : \mathcal{P_G} \rightarrow \mathcal{H}$.
 Then, the parametric grid in the physical domain $\mathcal{H}$ can be
    generated by the Gregory solid mapping.

 Suppose the parametric domain $\mathcal{P_G}$ has $n$ corners.
 In order to produce the parametric grid,
    the parametric domain $\mathcal{P_G}$ is first segmented into $n$ hexahedra using the dual operation (Algorithm 1),
    each hexahedron for a corner (Fig.~\ref{subfig:polyhedron}).
 Then, each hexahedron is uniformly discretized into a $M \times N \times L$ grid (Fig.~\ref{subfig:grid}).
 For conformity, on the common face of two adjacent hexahedra,
    the discretization is taken the same manner,
    resulting in the same grid.
 Finally, the grids in the parametric domain $\mathcal{P_G}$ is mapped into
    the physical domain $\mathcal{H}$ by the Gregory mapping
    $\mathcal{G}(p) : \mathcal{P_G} \rightarrow \mathcal{H}$,
    thus generating the parametric grid in the physical domain (Fig.~\ref{subfig:seperatedhead}).

\begin{algorithm}[!htp]
	\caption{Dual operation (refer to Fig.~\ref{fig:multi grid})}
	\begin{algorithmic}[1]
         \STATE Calculate the middle point of every edge of the polyhedral
                parametric domain $\mathcal{P_G}$;
         \STATE Calculate the barycenter of every face of $\mathcal{P_G}$;
         \STATE Calculate the barycenter of $\mathcal{P_G}$;
         \STATE Construct a hexahedron for each corner $c_l$ of
                $\mathcal{P_G}$,
                by linking the eight points,
                i.e., the middle points of the three edges adjacent to $c_l$, the barycenters of the three faces adjacent to $c_l$,
                the barycenter of $\mathcal{P_G}$,
                and the corner $c_l$ (refer to Fig.~\ref{subfig:hex},
                and \ref{subfig:polyhedron}).
	\end{algorithmic}
\end{algorithm}



 \section{Optimization}
 \label{sec:optimization}

 In this section, we develop a sparse optimization model to improve the
    algebraic quality, i.e., the quality of the parametrization, of the Gregory solid constructed in Section~\ref{sec:gregory_solid}.
 In order to employ the sparse optimization technique,
    the formulation of the objective function is based on the parametric grid generated using the method developed in Section~\ref{subs:parametric_grid}.

 It is well known that, a trivariate solid is valid in isogeometric analysis
    if its Jacobian is positive at any point.
 Note that the parametric grid in the physical domain $\mathcal{H}$ is a
    hexahedral mesh (Fig.~\ref{subfig:seperatedhead}),
    and the Jacobian of a hexahedral mesh is defined for each vertex of each hexahedron.
 Specifically, give the $s^{th} (s=0,1,\cdots,K)$ hexahedron of the
    hexahedral mesh with vertices
    $P_h = (x_h,y_h,z_h),\ h=0,1,\cdots,7$,
    and suppose
    $P_i = (x_i,y_i,z_i),\ P_j = (x_j,y_j,z_j),\ P_k(x_k,y_k,z_k),$
    are the three vertices adjacent to the vertex $P_h = (x_h,y_h,z_h)$.
 The order of the vertices of the tetrahedron $P_h - P_i P_j P_k$ is arranged
    in some specified orientation (clockwise or counterclockwise),
    so that the majority of Jacobians are positive.
 The scaled Jacobian at the vertex $P_h,\ h=0,1,\cdots,7$ of the $s^{th}$
    hexahedron is defined as~\cite{Knupp2015Achieving},
    \begin{equation*}
    J^s_h= det\left(\begin{array}{cccc}
    \frac{x_i - x_h}{\norm{\overrightarrow{P_i P_h}}_2} &
    \frac{x_j - x_h}{\norm{\overrightarrow{P_j P_h}}_2} &
    \frac{x_k - x_h}{\norm{\overrightarrow{P_k P_h}}_2}  \\
    \frac{y_i - y_h}{\norm{\overrightarrow{P_i P_h}}_2} &
    \frac{y_j - y_h}{\norm{\overrightarrow{P_j P_h}}_2} &
    \frac{y_k - y_h}{\norm{\overrightarrow{P_k P_h}}_2}  \\
    \frac{z_i - z_h}{\norm{\overrightarrow{P_i P_h}}_2} &
    \frac{z_j - z_h}{\norm{\overrightarrow{P_j P_h}}_2} &
    \frac{z_k - z_h}{\norm{\overrightarrow{P_k P_h}}_2}
    \end{array}\right),\
    h = 0,1,\cdots,7,\ s=0,1,\cdots, K,
    \end{equation*}
    where $\norm{\cdot}_2$ is the 2-norm of a vector.
 Thus, the scaled Jacobians for the parametric grid of the physical domain
    $\mathcal{H}$ can be organized in a vector $Jac$,
    \begin{equation}
    \label{eq:jac_vector}
    Jac = [J^0_0, J^0_1,\cdots,J^0_7, \cdots, J^K_0, J^K_1, \cdots, J^K_7].
    \end{equation}
 Moreover, by defining two functions,
 \begin{equation*}
    \begin{array}{ccc}
        pos(J^s_h) =
        \begin{cases}
            J^s_h, & J^s_h \geq 0, \\
            0      & J^s_h < 0,
        \end{cases}
    &
    \qquad
    \text{and}
    \qquad
    &
        neg(J^s_h) =
        \begin{cases}
            0,      & J^s_h > 0, \\
            J^s_h   & J^s_h \leq 0,
        \end{cases}
    \end{array}
 \end{equation*}
 the Jacobian vector $Jac$~\pref{eq:jac_vector} can be decomposed into two parts, i.e.,
 \begin{equation*}
 Jac = Jac^{+} + Jac^{-},
 \end{equation*}
 where,
 \begin{equation}
 \label{eq:pos_jac}
   Jac^{+} = [pos(J^0_0), pos(J^0_1),\cdots, pos(J^0_7), \cdots,
                pos(J^K_0), pos(J^K_1), \cdots, pos(J^K_7)]
 \end{equation}
    contains the positive and zero elements of the vector $Jac$~\pref{eq:jac_vector},
    and
    \begin{equation}
    \label{eq:neg_jac}
      Jac^{-} = [neg(J^0_0), neg(J^0_1),\cdots, neg(J^0_7), \cdots,
                neg(J^K_0), neg(J^K_1), \cdots, neg(J^K_7)]
   \end{equation}
    contains the negative and zero elements of the vector $Jac$~\pref{eq:jac_vector}.

 On one hand, to improve the validity of the Gregory solid,
    two objective functions are required.
 First, the less the number of the vertices with negative Jacobian,
    the better the validity,
    which is formulated as the sparse optimization objective function,
    \begin{equation}
    \label{eq:sparse_obj}
        E_{sparse} = \norm{Jac^-}_0,
    \end{equation}
    where $\norm{Jac^-}_0$ is the 0-norm of $Jac^-$,
    that is, the number of nonzero elements of the vector $Jac^-$.
 Second, the larger the sum of the positive Jacobians,
    the better the validity,
    which is modelled as the following objective function (refer to~\pref{eq:jac_vector}),
    \begin{equation}
    \label{eq:positive_obj}
        E_{positive} = \sum_{J^s_h \geq 0} \frac{1}{J^s_h+\epsilon}.
    \end{equation}
 Experiments show that $\epsilon = 10^{-5}$ can lead to desirable results.

 On the other hand, to improve the smoothness of the parametric grid of the
    physical domain $\mathcal{H}$,
    the Laplace smoothing function is taken as an objective function,
    \begin{equation}
    \label{eq:laplace}
        E_{smooth} = \sum_i \norm{ P_i -
                     \frac{1}{\# N(P_i)}
                     \sum_{P_j \in N(P_i)} P_j }_2^2,
    \end{equation}
    where $P_i$ is a vertex of the parametric grid,
    $N(P_i)$ denotes the set of one-ring adjacent vertices to $P_i$,
    and $\# N(P_i)$ is the number of the one-ring adjacent vertices to $P_i$.

 Consequently, the whole objective function $E$ is taken as a linear
    combination of the three aforementioned objective functions,
    \begin{equation*}
        E = E_{smooth} + \mu E_{positive} + \nu E_{sparse},
    \end{equation*}
    where the weights $\mu > 0$ and $\nu > 0$ are utilized to balance the three items.
 Returning to the Gregory solid representation~\pref{eq:gregory_solid},
    we can see that the variables that can be adjusted for improving the parametric quality are the control points of the three bi-cubic B-spline vector functions
    $T^{top}(u,v)$, $T^{lft}(u,w)$, and $T^{rgt}(v,w)$ \pref{eq:vector_functions}.
 Denoting the set of these control points as $X$,
    the whole optimization problem can be formulated as,
 \begin{equation}
 \label{eq:org_optimization_problem}
 \begin{split}
    \min_{X}
    \sum_i
    \norm{P_i - \frac{1}{\#N(P_i)}\sum_{P_j \in N(P_i)}P_j}_2^2
    +
    \mu \sum_{J^s_h \geq 0} \frac{1}{J^s_h+\epsilon}
    +
    \nu \norm{Jac^-}_0.
 \end{split}
 \end{equation}

 In fact, for the convenience of the optimization problem
    \pref{eq:org_optimization_problem} to be solved by the alternating direction method of multipliers (ADMM)~\cite{Boyd2010Distributed},
    the $0-$norm of the sparse item is replaced by the $1-$norm,
    and the optimization problem is changed to,
 \begin{equation}
 \label{eq:optimization_problem}
 \begin{split}
    &\min_{X, Y, Z}
    \sum_i
    \norm{P_i(X) - \frac{1}{\#N(P_i)}\sum_{P_j \in N(P_i)}P_j (X)}_2^2
    +
    \mu \sum_{J^s_h \geq 0} \frac{1}{J^s_h(Y)+\epsilon}
    +
    \nu \norm{Jac^-(Z)}_1 \\
    & s.t. \quad X = Y = Z.
 \end{split}
 \end{equation}
 Then, we can develop the format of ADMM for solving the above optimization
    problem~\pref{eq:optimization_problem},
 \begin{align}
 X^{t+1} &\leftarrow \arg\min_X\left(\sum_{i}\norm{P_i(X)-\frac{1}{\# N(P_i)}\sum\limits_{P_j \in N(P_i)}P_j(X)}_2^2+\frac{\rho}{2}
 \norm{
 \begin{bmatrix} X \\ X \end{bmatrix}-\begin{bmatrix} Y^t \\ Z^t \end{bmatrix}+\begin{bmatrix} U_Y^t \\ U_Z^t \end{bmatrix}}^2_2 \right), \label{eq:first_op} \\
 Y^{t+1} &\leftarrow \arg\min_Y\left(\mu\sum_{J^s_h \geq 0} \frac{1}{J^s_h(Y)+\epsilon}+\frac{\rho}{2}\|X^{t+1}-Y+U_Y^t\|^2_2\right),
  \label{eq:second_op} \\\
 Z^{t+1} &\leftarrow \arg\min_Z\left(\nu\norm{Jac^-(Z)}_1+\frac{\rho}{2}\norm{X^{t+1}-Z+U_Z^t}^2_2 \right), \label{eq:third_op} \\
 U^{t+1} &\leftarrow U^t+\rho\left(\begin{bmatrix} X^{t+1} \\ X^{t+1} \end{bmatrix}-\begin{bmatrix} Y^{t+1} \\ Z^{t+1} \end{bmatrix}\right),\\
         & t = 0,1,2,\cdots
 \end{align}
 where the factor $\rho$ is a penalty parameter,
    and $\rho = 1$ is a desirable selection for fast convergence.
 The initial values $X^0 = Y^0 = Z^0$ are constructed by the method presented
    in Remark~\ref{rmk:initial_patch},
    and we set $U^0 = 0$.
 In the ADMM format developed above,
    each individual updating step is a small optimization itself,
    and can be solved efficiently.
 Specifically, in our implementation,
    the gradient descent method~\cite{Avriel1976Nonlinear} is employed to solve the first
    optimization problem~\pref{eq:first_op},
    and the sub-gradient descent method~\cite{Avriel1976Nonlinear} is adopted to solve the second~\pref{eq:second_op} and the third optimization problem~\pref{eq:third_op}.


\begin{figure}[!htb]
  \begin{center}
  \subfigure[]{
    \label{subfig:duckmodel}
    \includegraphics[width=0.20\textwidth]{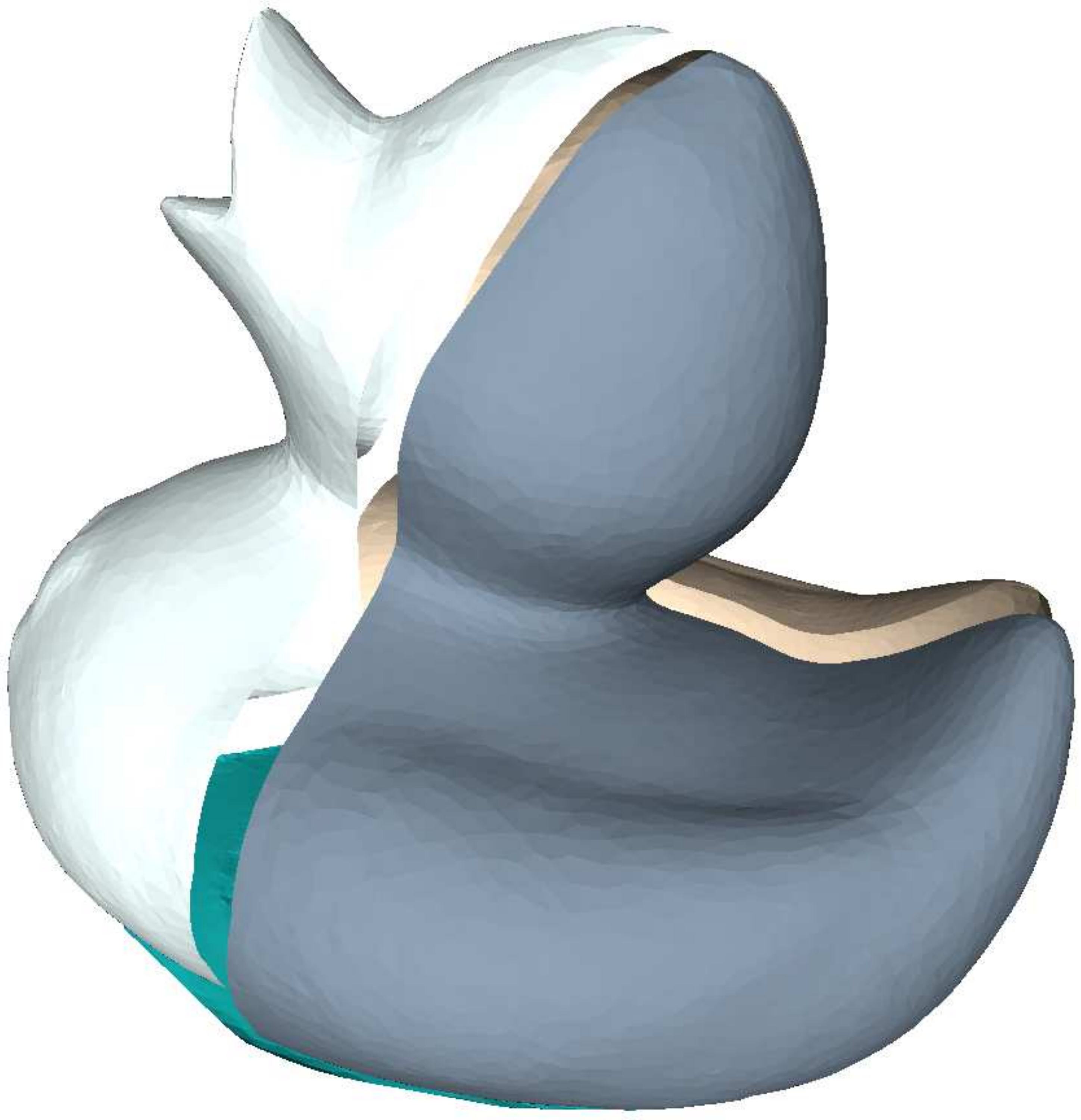}}
  \subfigure[]{
    \label{subfig:balljointmodel}
    \includegraphics[width=0.18\textwidth]{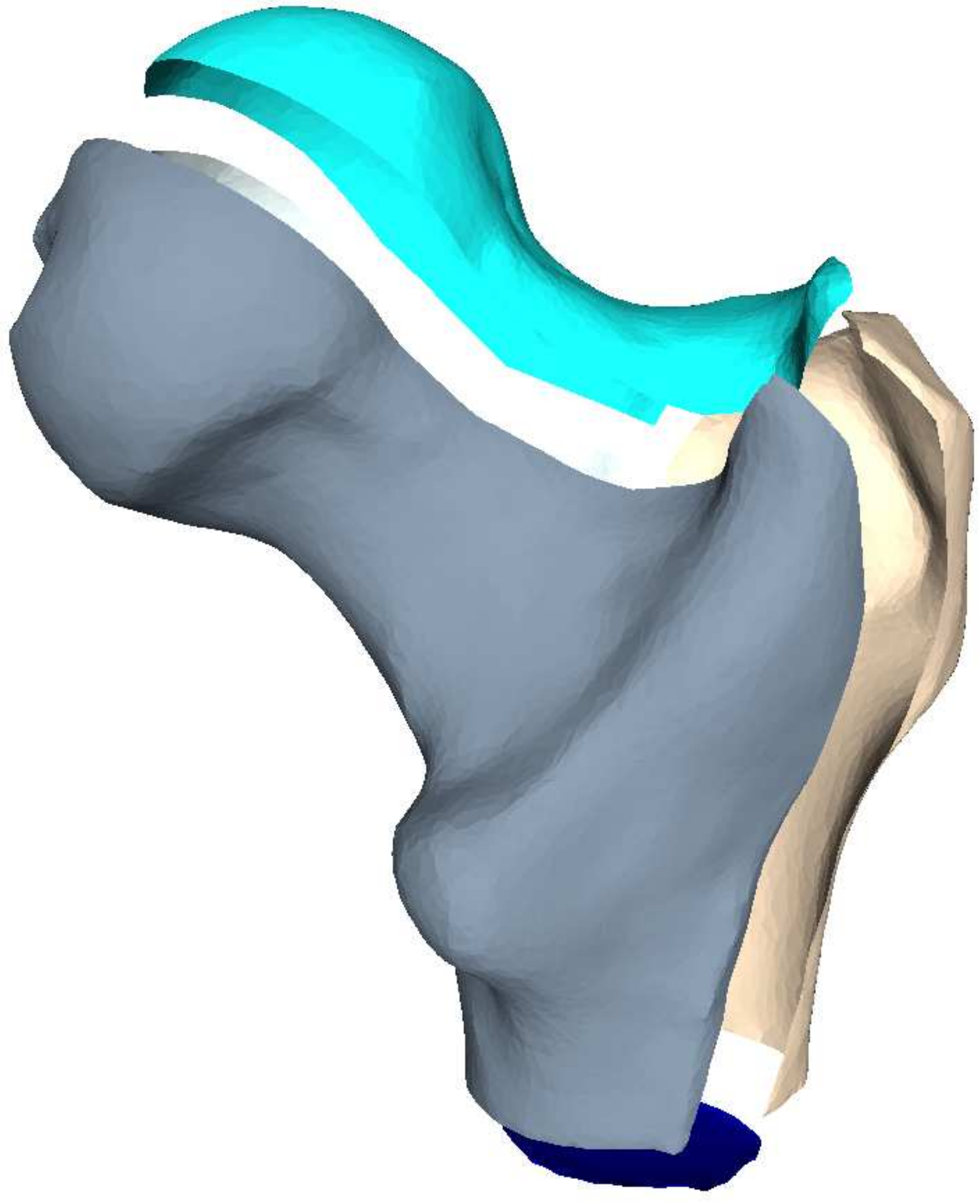}}
  \subfigure[]{
    \label{subfig:toothmodel}
    \includegraphics[width=0.19\textwidth]{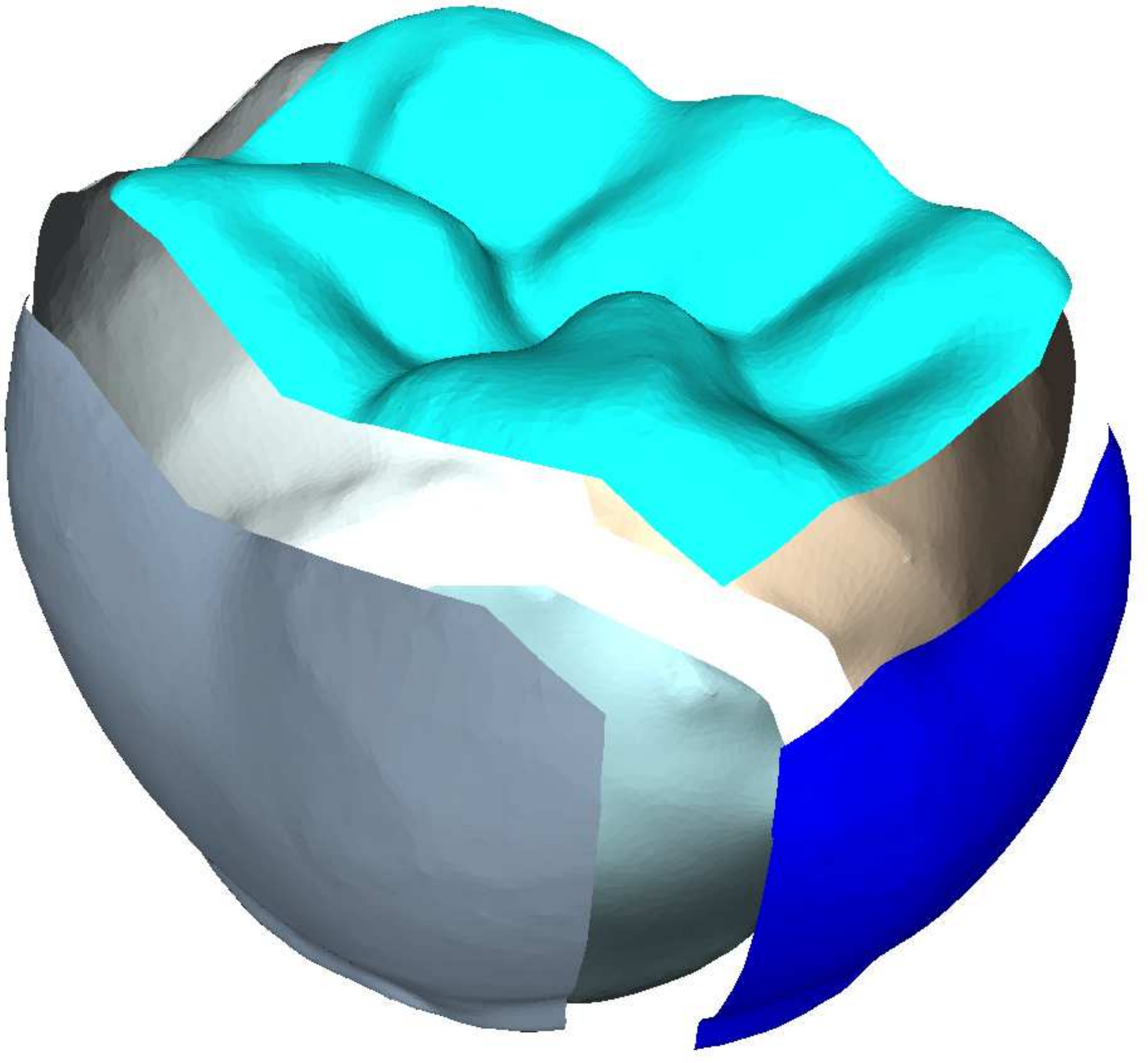}}
  \subfigure[]{
    \label{subfig:headmodel}
    \includegraphics[width=0.17\textwidth]{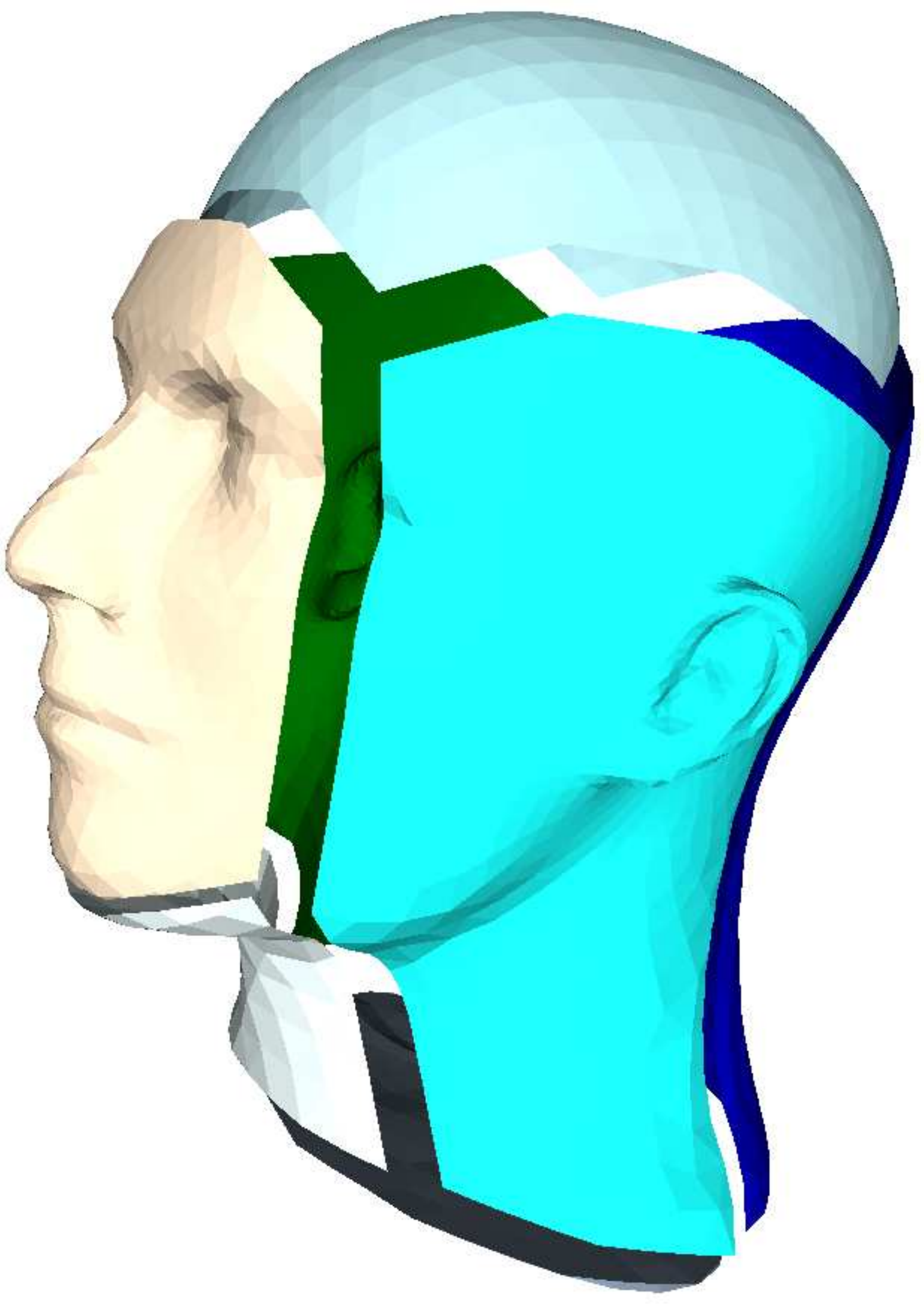}}
  \subfigure[]{
    \label{subfig:moaimodel}
    \includegraphics[width=0.13\textwidth]{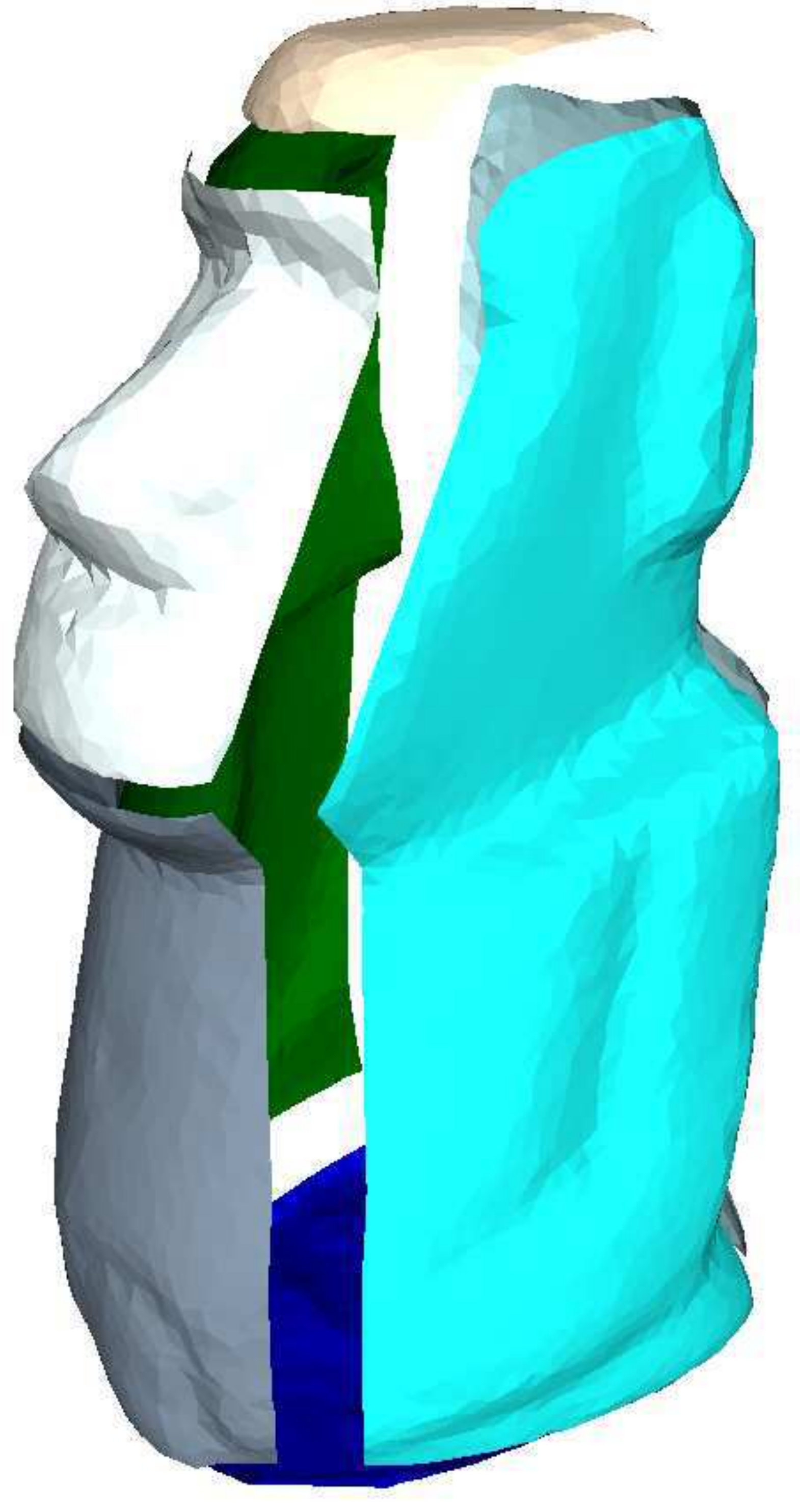}}
  \subfigure[]{
    \label{subfig:duckgrid}
    \includegraphics[width=0.18\textwidth]{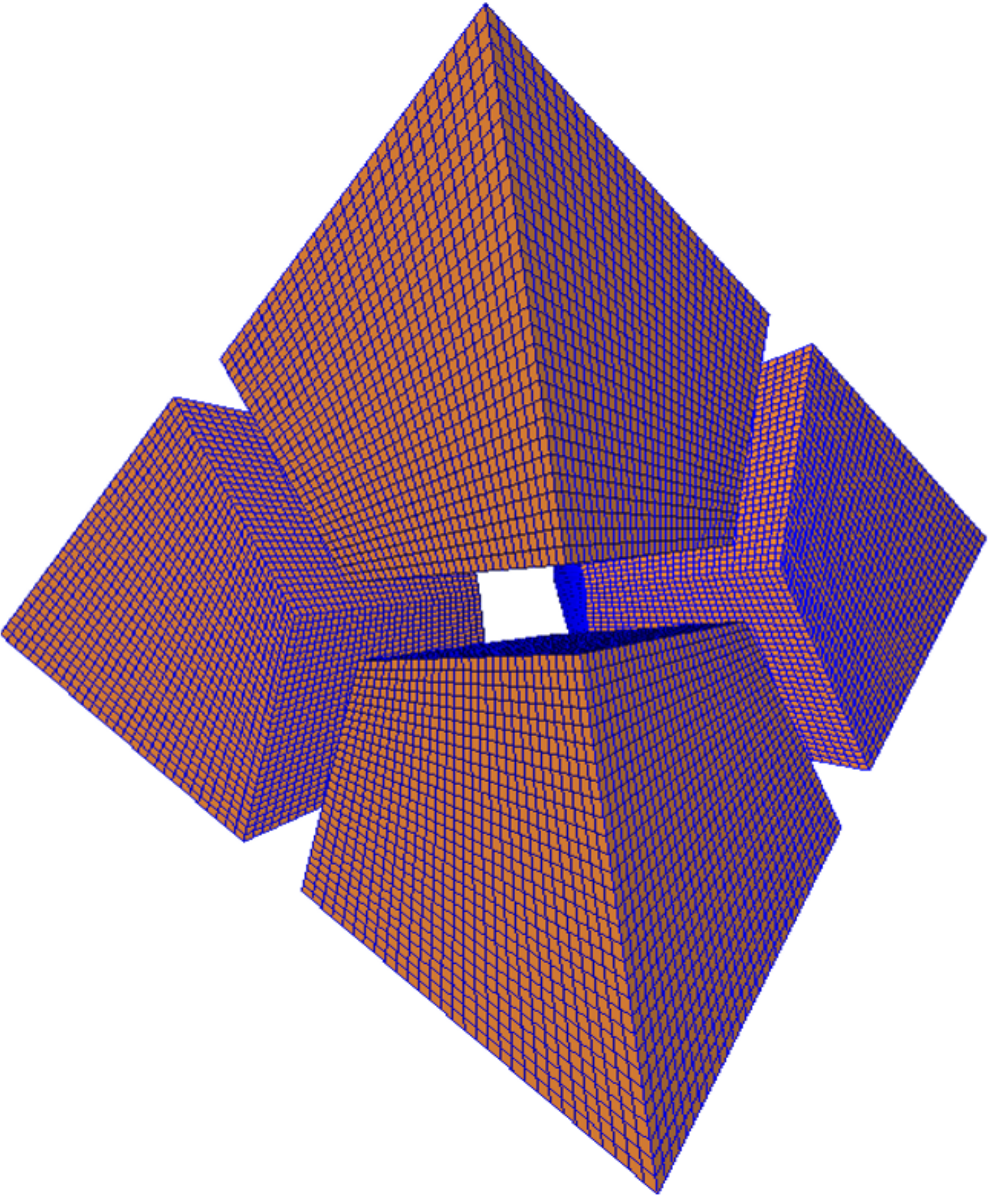}}
  \subfigure[]{
    \label{subfig:balljointgrid}
    \includegraphics[width=0.18\textwidth]{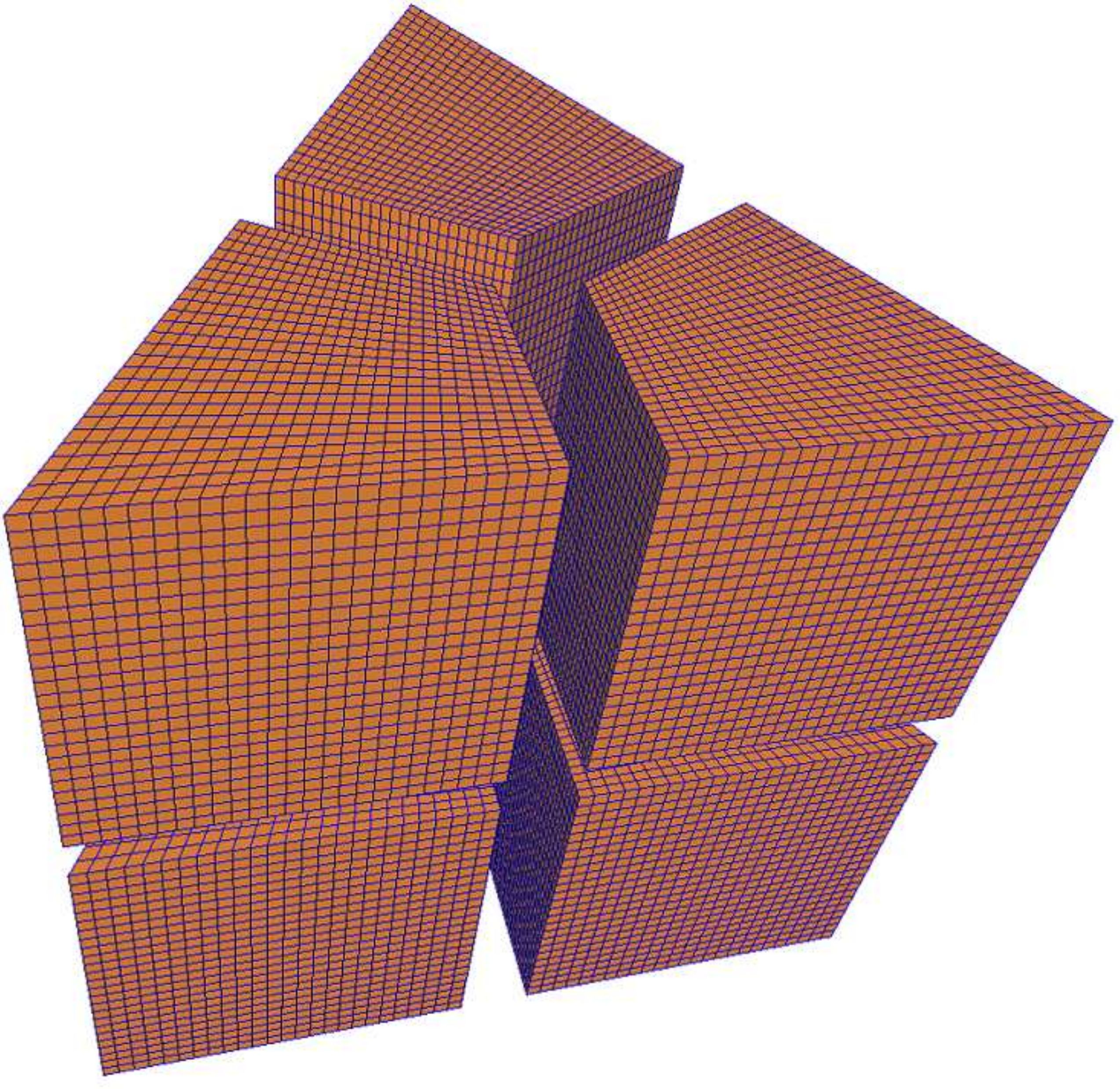}}
  \subfigure[]{
    \label{subfig:toothgrid}
    \includegraphics[width=0.18\textwidth]{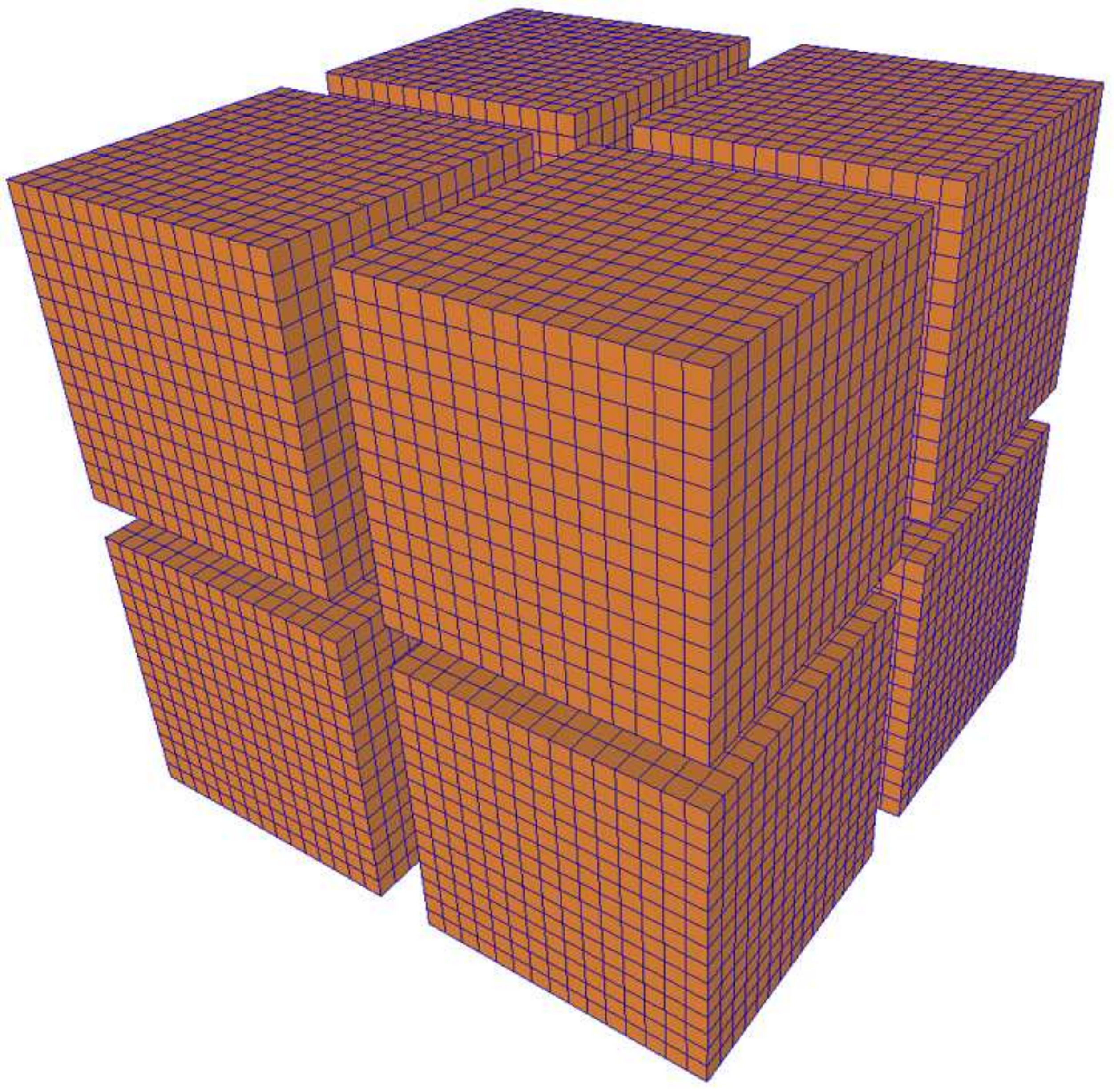}}
  \subfigure[]{
    \label{subfig:headgrid}
    \includegraphics[width=0.18\textwidth]{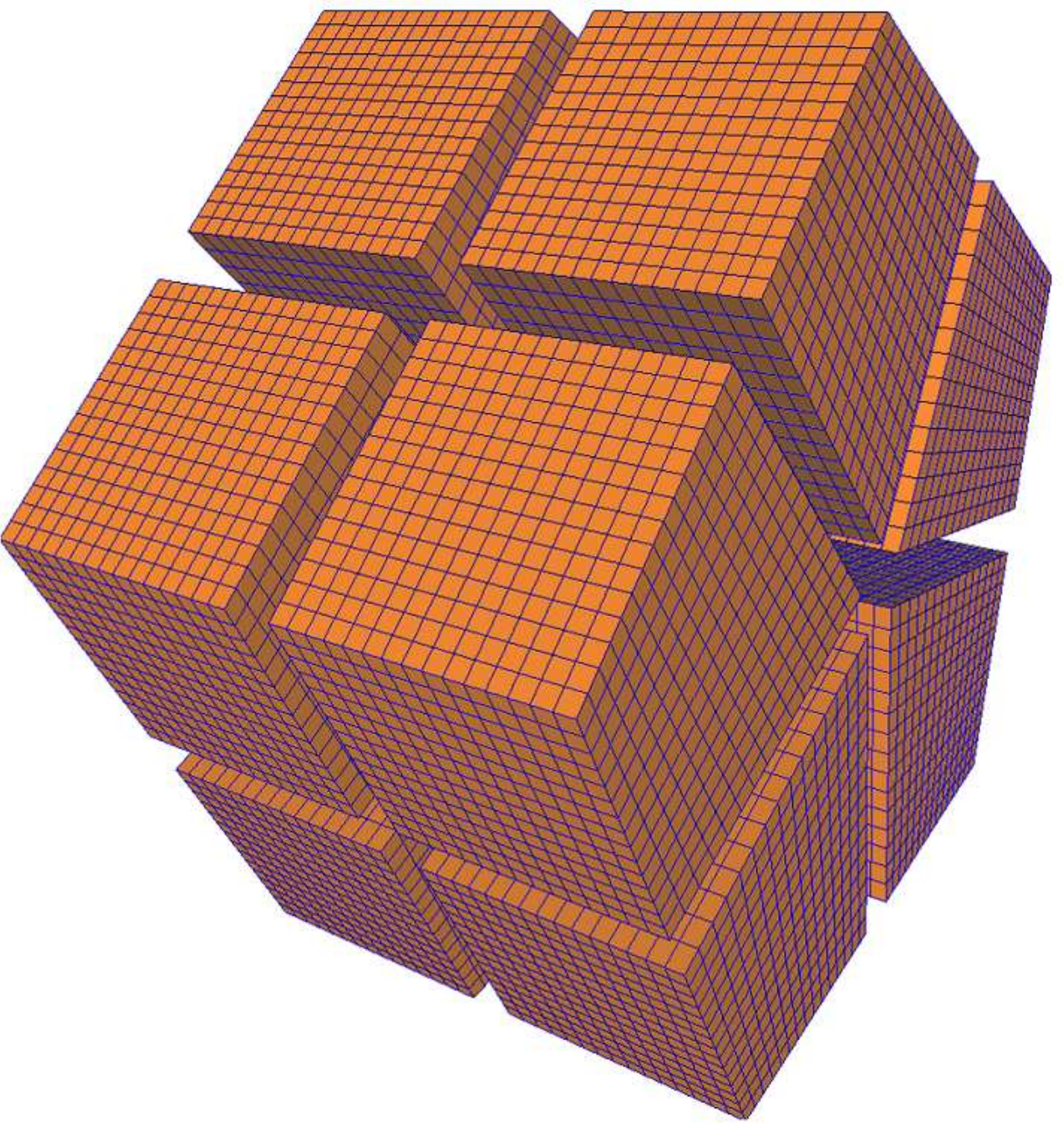}}
  \subfigure[]{
    \label{subfig:moaigrid}
    \includegraphics[width=0.18\textwidth]{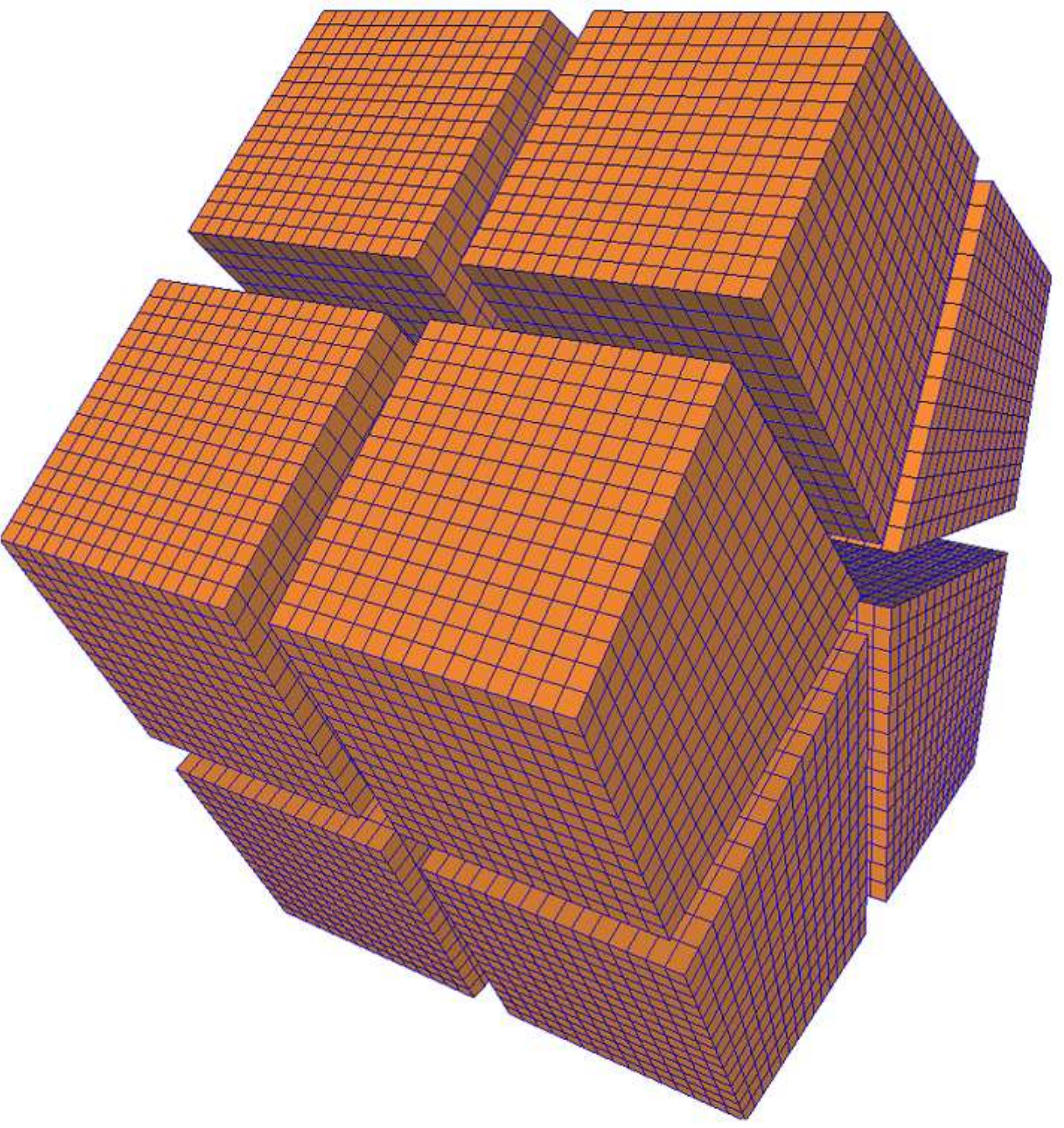}}
      \caption{\small The input mesh models, \emph{Duck}(a), \emph{Ball joint}(b), \emph{Tooth}(c), \emph{Head}(d), and \emph{Moai}(e);
      their separated parametrization domain: tetrahedron (f), pentahedron (triangular prism) (g), hexahedron (h), heptahedron (pentagonal prism)(i), and heptahedron (pentagonal prism)(j).}
   \label{fig:triangular_mesh}
  \end{center}
\end{figure}

 \section{Results and discussions}
 \label{sec:results}

 In this section,
    some experimental results are illustrated to demonstrate the effectiveness and efficiency of the Gregory solid construction and optimization algorithm developed above.
 All of the experimental examples are run on a PC with 3.60GHz CPU i7-4790
    and 16G memory.
 As stated in Section~\ref{sec:gregory_solid},
    the input to our algorithm is a polyhedral hollowed physical domain $\mathcal{H}$,
    where each corner is adjacent to just three boundary patches.
 Although the boundary patches can be any type of parametric surfaces,
    in our experiments, triangular meshes are taken as the boundary patches.
 Moreover, we employ the conformal parametrization method
    \cite{Floater1997Parametrization} to calculate their parametrization.
 Thus, the triangular meshes become parametric surfaces with their
    parametrization.

 In Fig.~\ref{fig:triangular_mesh},
    the input polyhedral physical domains and the parametric domains of the corresponding Gregory solids are illustrated.
 While the input physical domains are demonstrated in
    Figs.~\ref{fig:triangular_mesh}(a)-(e),
    the parametric domains of the corresponding Gregory solids are presented in Figs.~\ref{fig:triangular_mesh}(f)-(j).
 Specifically, the models illustrated in Fig.\ref{fig:triangular_mesh}(a)-(e)
    are \emph{Duck}, \emph{Ball Joint}, \emph{Tooth}, \emph{Head}, and \emph{Moai}, respectively,
    and their parametric domains are
    tetrahedron (Fig.~\ref{fig:triangular_mesh}(f)),
    pentahedron (triangular prism, Fig.~\ref{fig:triangular_mesh}(g)), hexahedron (Fig.~\ref{fig:triangular_mesh}(h)),
    heptahedron (pentagonal prism, Fig.~\ref{fig:triangular_mesh}(i)),
    and heptahedron (pentagonal prism, Fig.~\ref{fig:triangular_mesh}(j)),
    respectively.
 Additionally, for each input model,
    the numbers of mesh vertices, triangular faces, and boundary patches are listed in Table~\ref{tbl:stat}.
 It can be seen that,
    the numbers of mesh vertices range from 3928 to 15154,
    and the numbers of faces from 7852 to 30304.

  \begin{table*}[!htb]
  \centering
  \footnotesize
  \caption{Statistical data of the Gregory solids generation method developed in this paper.}
  \label{tbl:stat}
  \begin{threeparttable}
  \begin{tabular}{| c | c | c | c | c | c | c | c | c | c |}
  \hline
    {model}    & {\#vert.\tnote{1}} & {\#face\tnote{2}} & {\#boundary\tnote{3}} & {\#grid\tnote{4}} &{avg.J.\tnote{5}} & {min.J.\tnote{5}} & {max.J.\tnote{5}} & {$J^-/J$ \tnote{6}} & {time\tnote{7}}\\
  \hline
    Duck       & 10461  & 20918 & 4 & $ 30 \times 30 \times 30$ &0.8006 & -0.5763 &0.9995 & 0.176\% & 2386.22\\
  \hline
    Ball joint & 10936  & 21868 & 5 & $ 26 \times 26 \times 26$ &0.8009 & -0.1495 & 0.9991 & 0.071\% & 1820.22\\
  \hline
    Tooth      & 15154  & 30304 & 6 & $ 15 \times 15 \times 15$ &0.9362 & 0.0133 & 0.9999 & 0 & 130.22\\
  \hline
    Head       & 3928  & 7852 & 7 & $ 16 \times 16 \times 16$ &0.8428 & -0.4752 & 0.9997 & 0.172\% & 1219.87 \\
  \hline
    Moai       & 5685  & 11366 & 7 & $ 18 \times 18 \times 18$ &0.8881 & -0.4693 & 0.9999 & 0.014\% & 1180.88\\
  \hline
  \end{tabular}
 \begin{tablenotes}
    \item[1] Number of vertices of the input triangular mesh model.
    \item[2] Number of faces of the input triangular mesh model.
    \item[3] Number of boundary patches of the physical domain.
    \item[4] Resolution of the discretized grid in each segmented parametric sub-domain.
    \item[5] Average, minimal, and maximal scaled Jacobian values.
    \item[6] Ratio between the volume of region with negative Jacobian and the volume of the whole model.
    \item[7] Running time in seconds.
 \end{tablenotes}
 \end{threeparttable}
 \end{table*}

\begin{figure}[!htb]
  \begin{center}
  \subfigure[]{
    \label{subfig:duckhex}
    \includegraphics[width=0.22\textwidth]{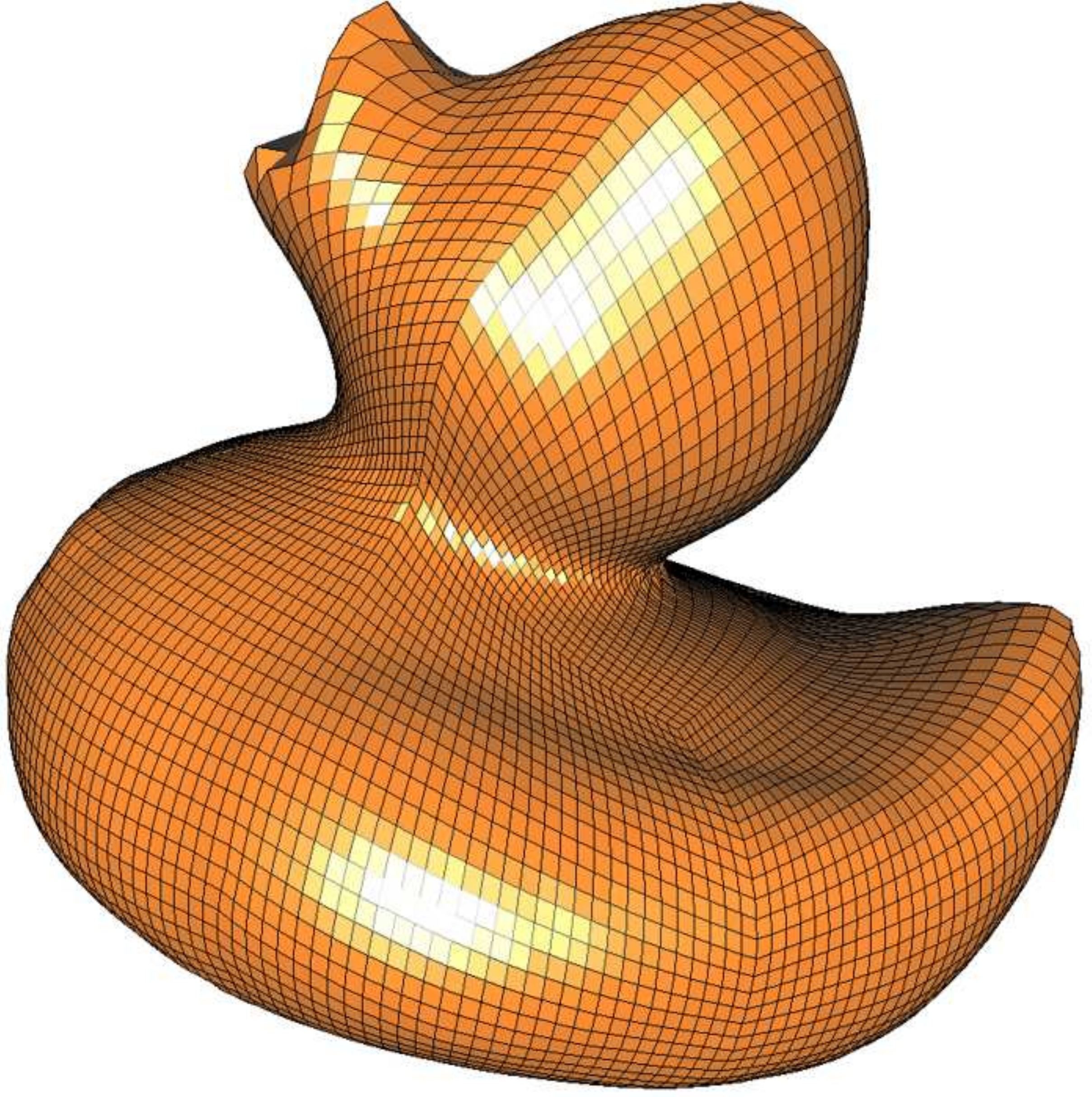}}
  \subfigure[]{
    \label{subfig:balljointhex}
    \includegraphics[width=0.18\textwidth]{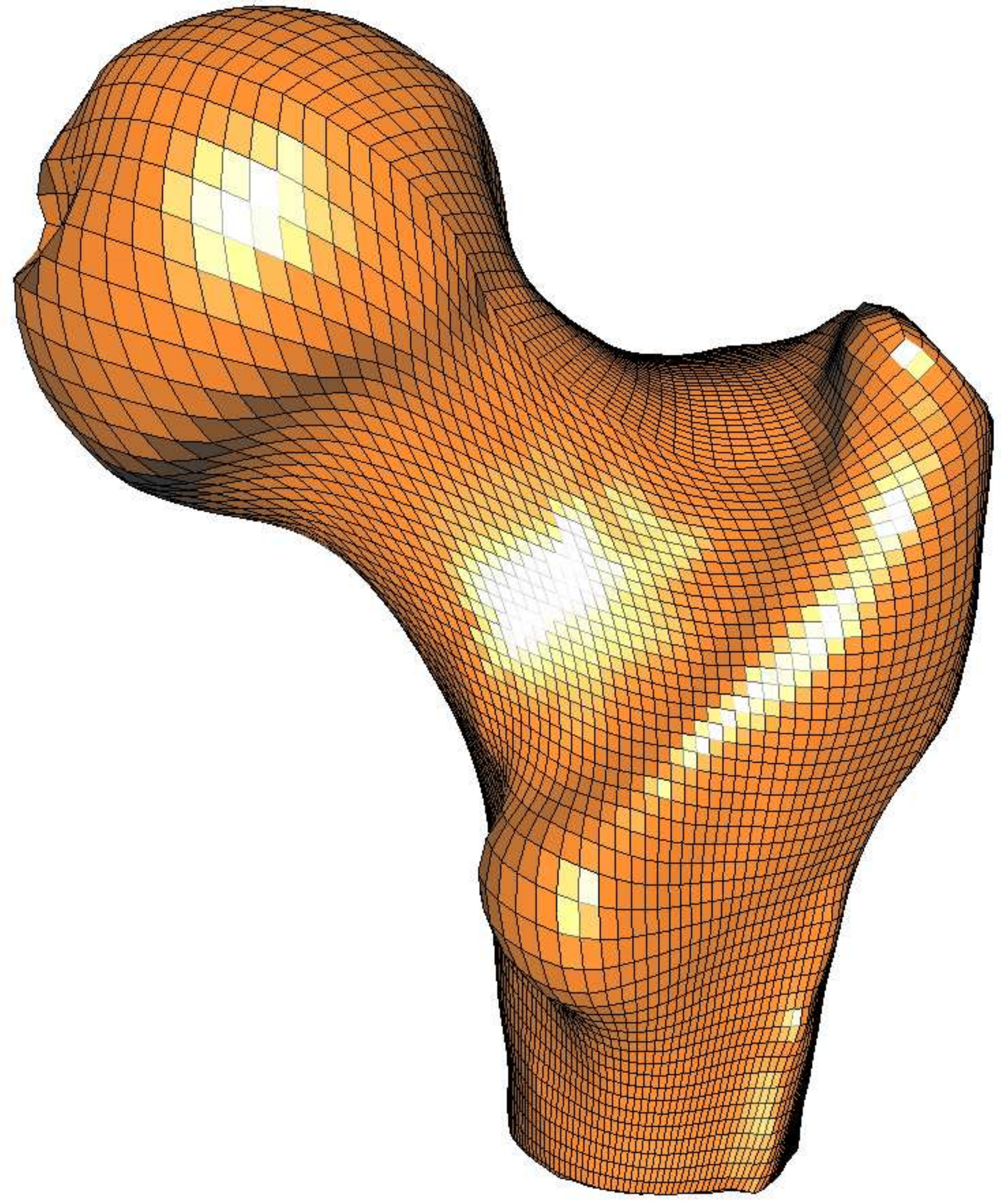}}
  \subfigure[]{
    \label{subfig:toothhex}
    \includegraphics[width=0.20\textwidth]{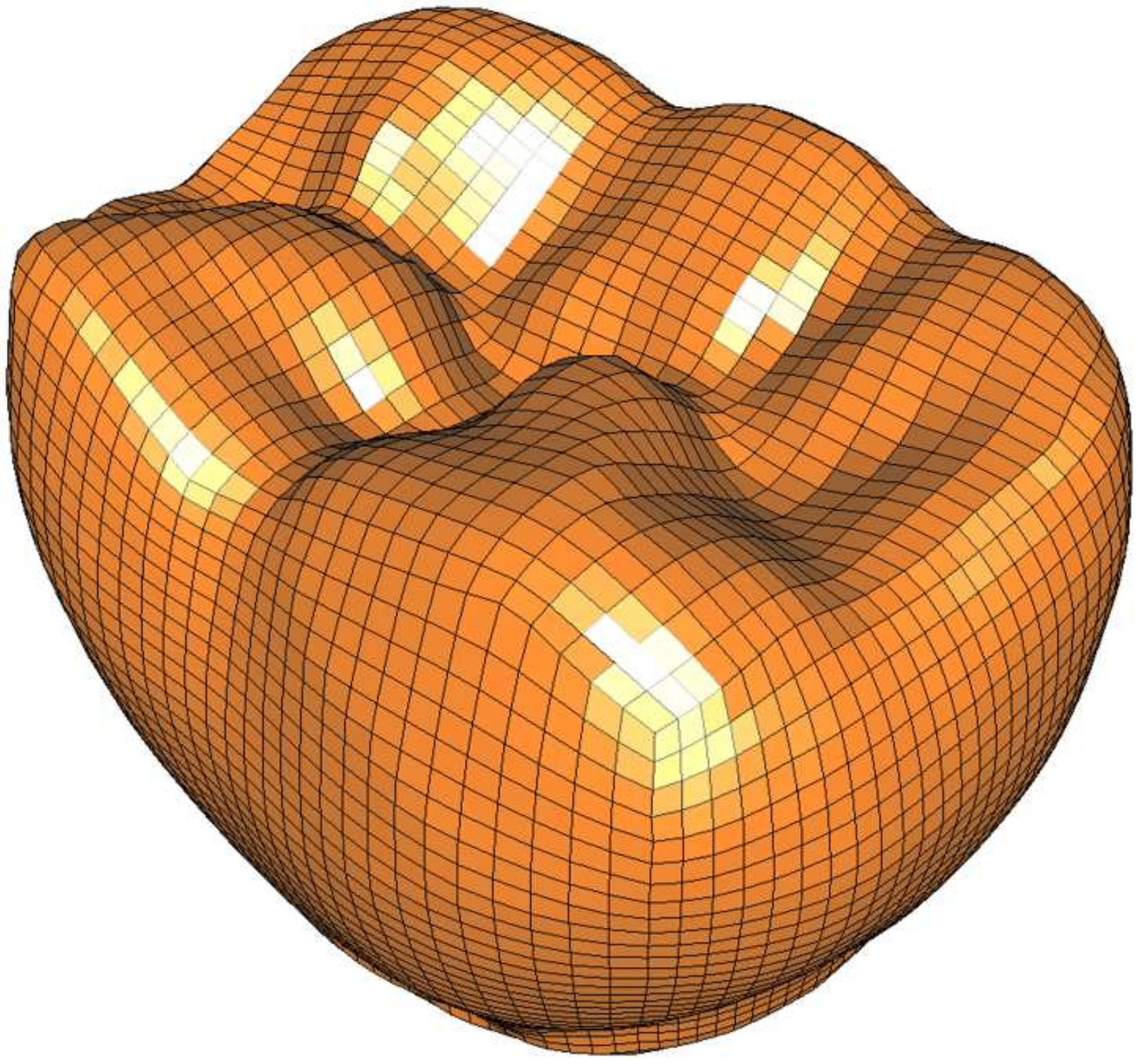}}
  \subfigure[]{
    \label{subfig:headhex}
    \includegraphics[width=0.17\textwidth]{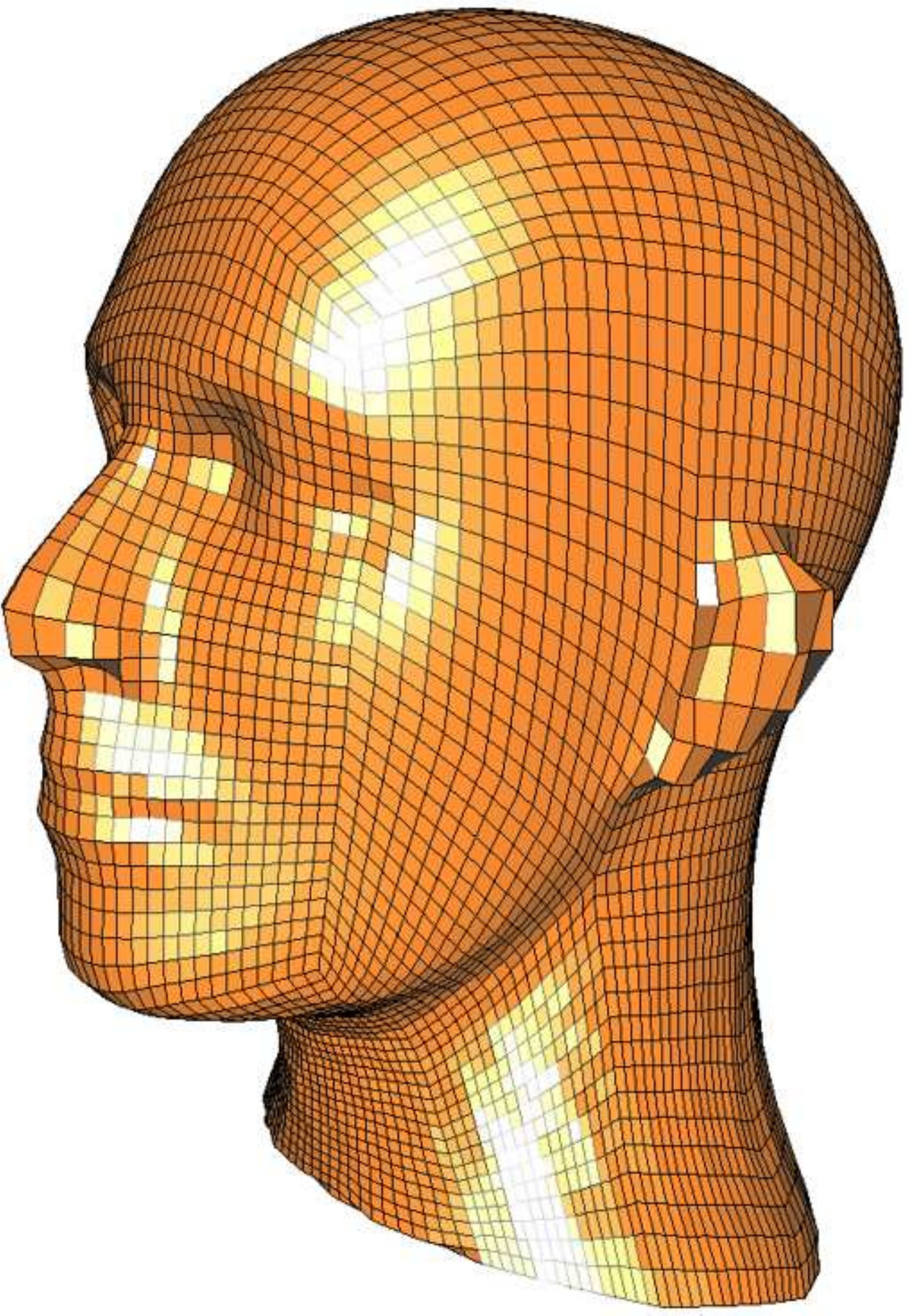}}
  \subfigure[]{
    \label{subfig:moaihex}
    \includegraphics[width=0.14\textwidth]{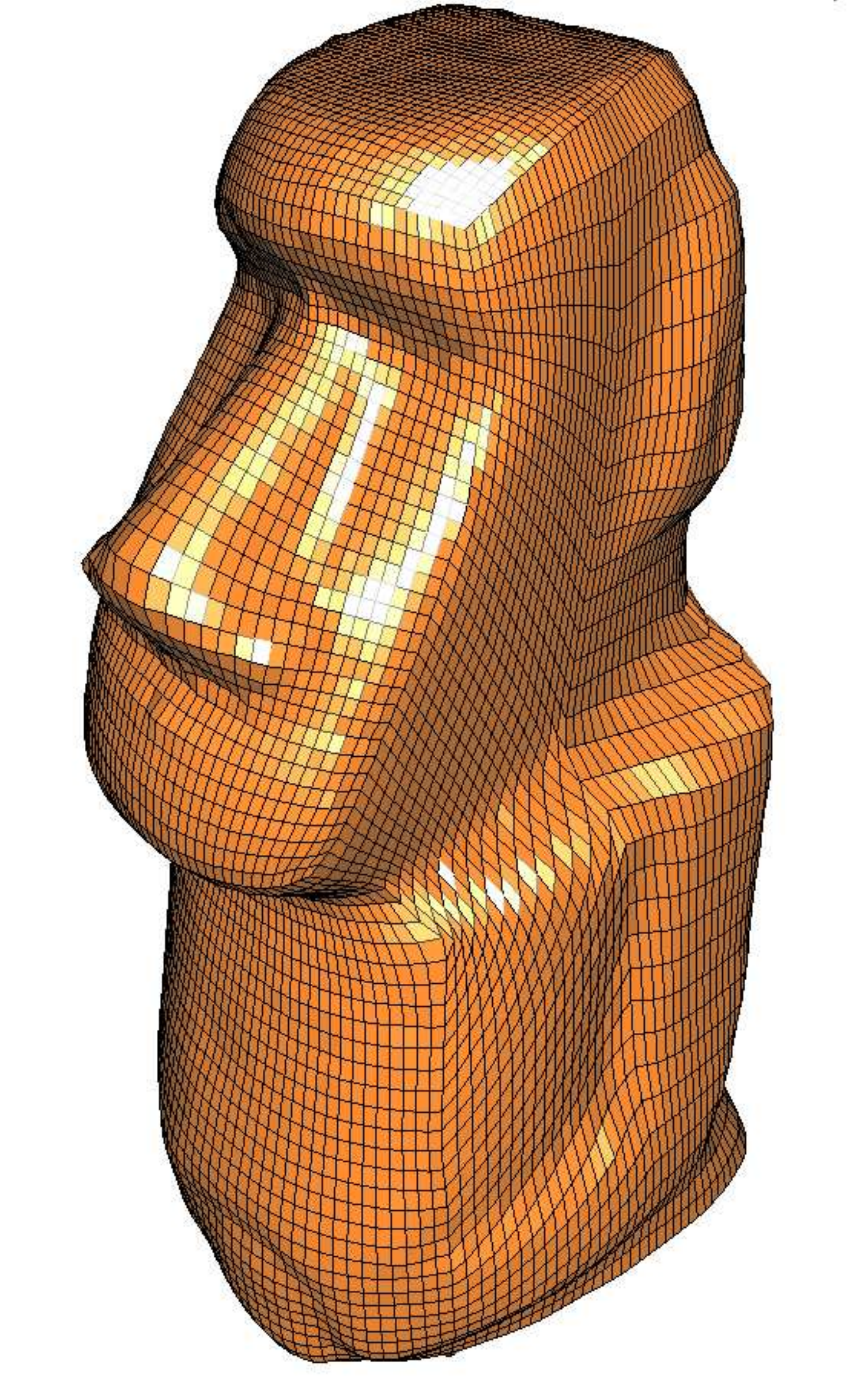}}
  \subfigure[]{
    \label{subfig:duckcut}
    \includegraphics[width=0.22\textwidth]{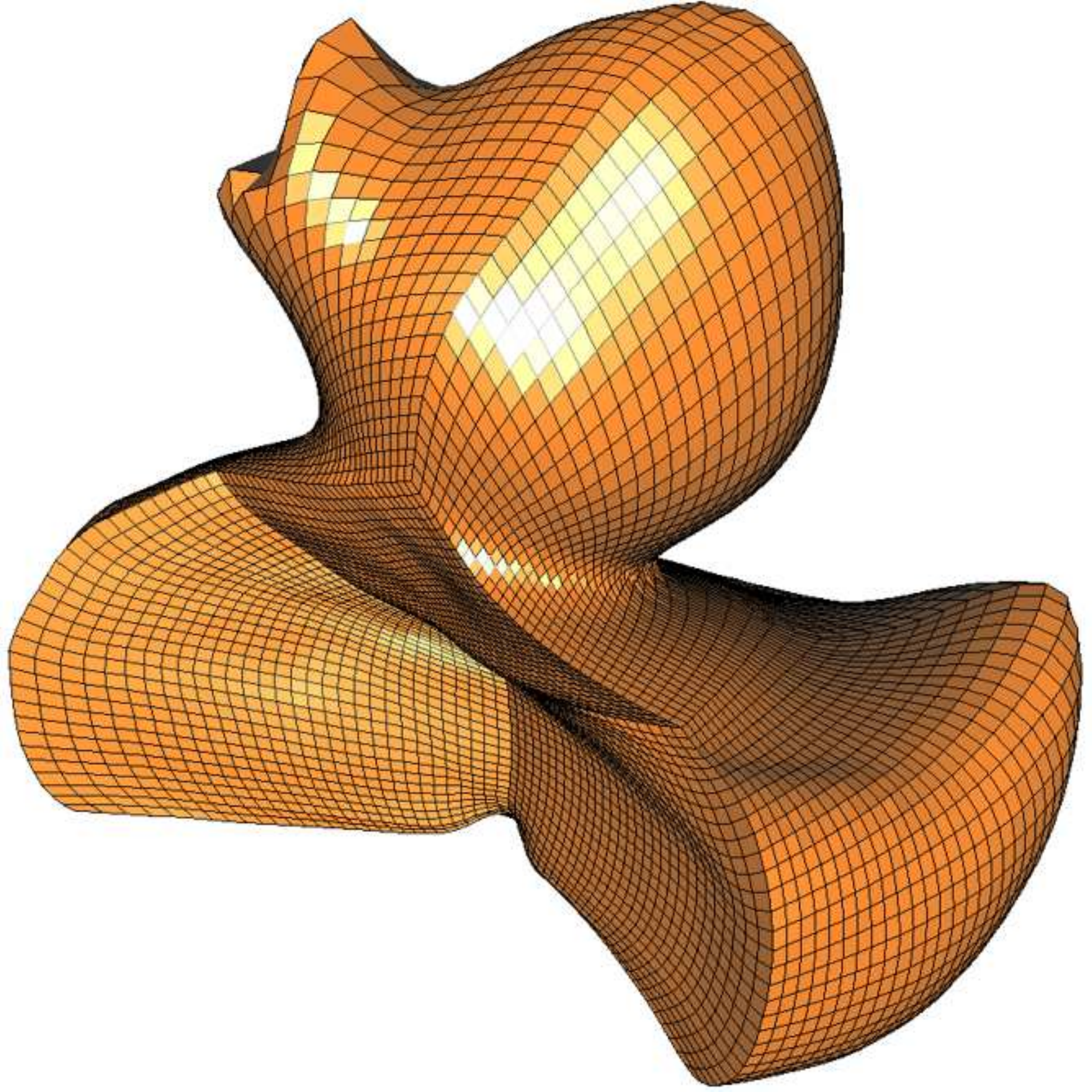}}
  \subfigure[]{
    \label{subfig:balljointcut}
    \includegraphics[width=0.18\textwidth]{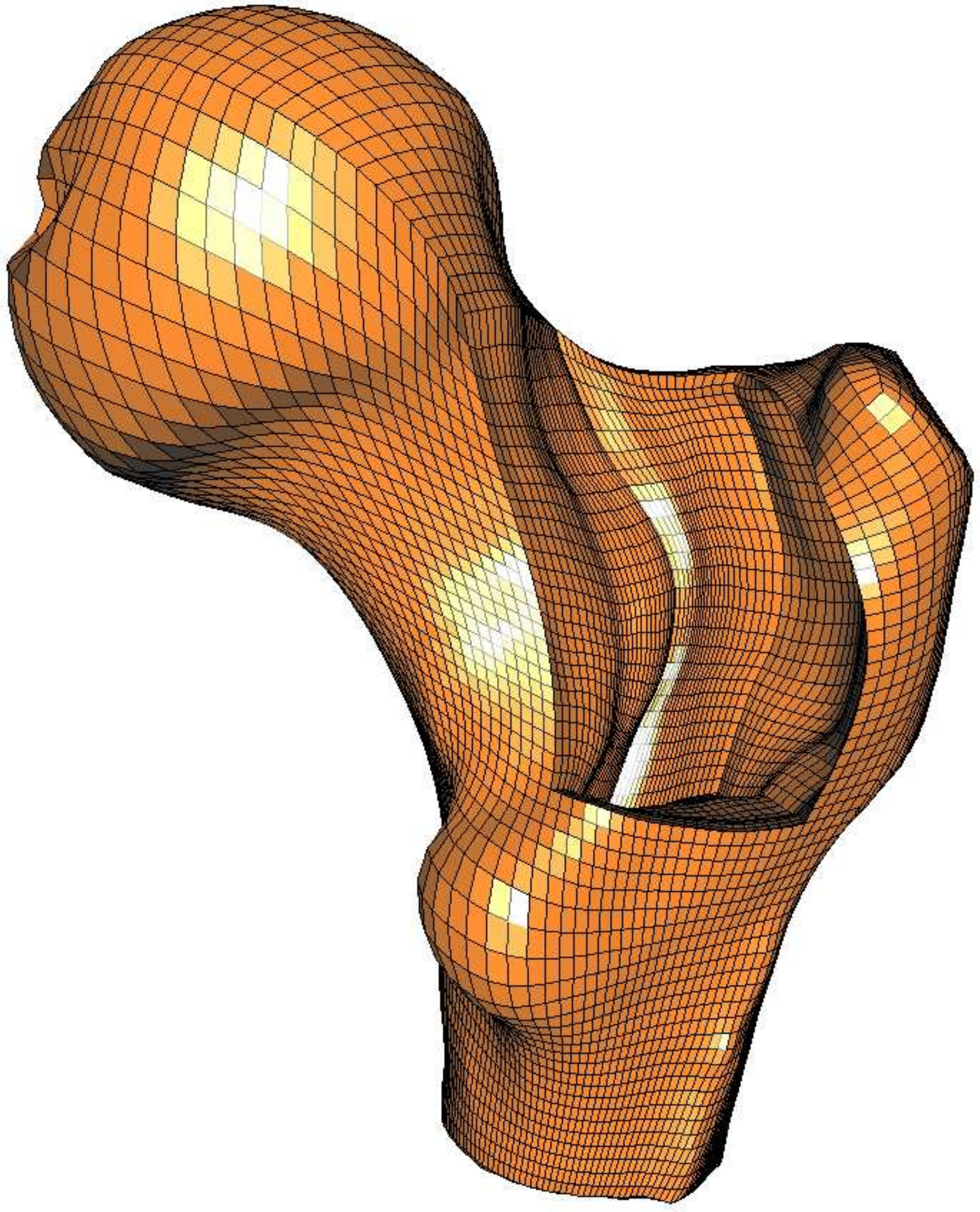}}
  \subfigure[]{
    \label{subfig:toothcut}
    \includegraphics[width=0.20\textwidth]{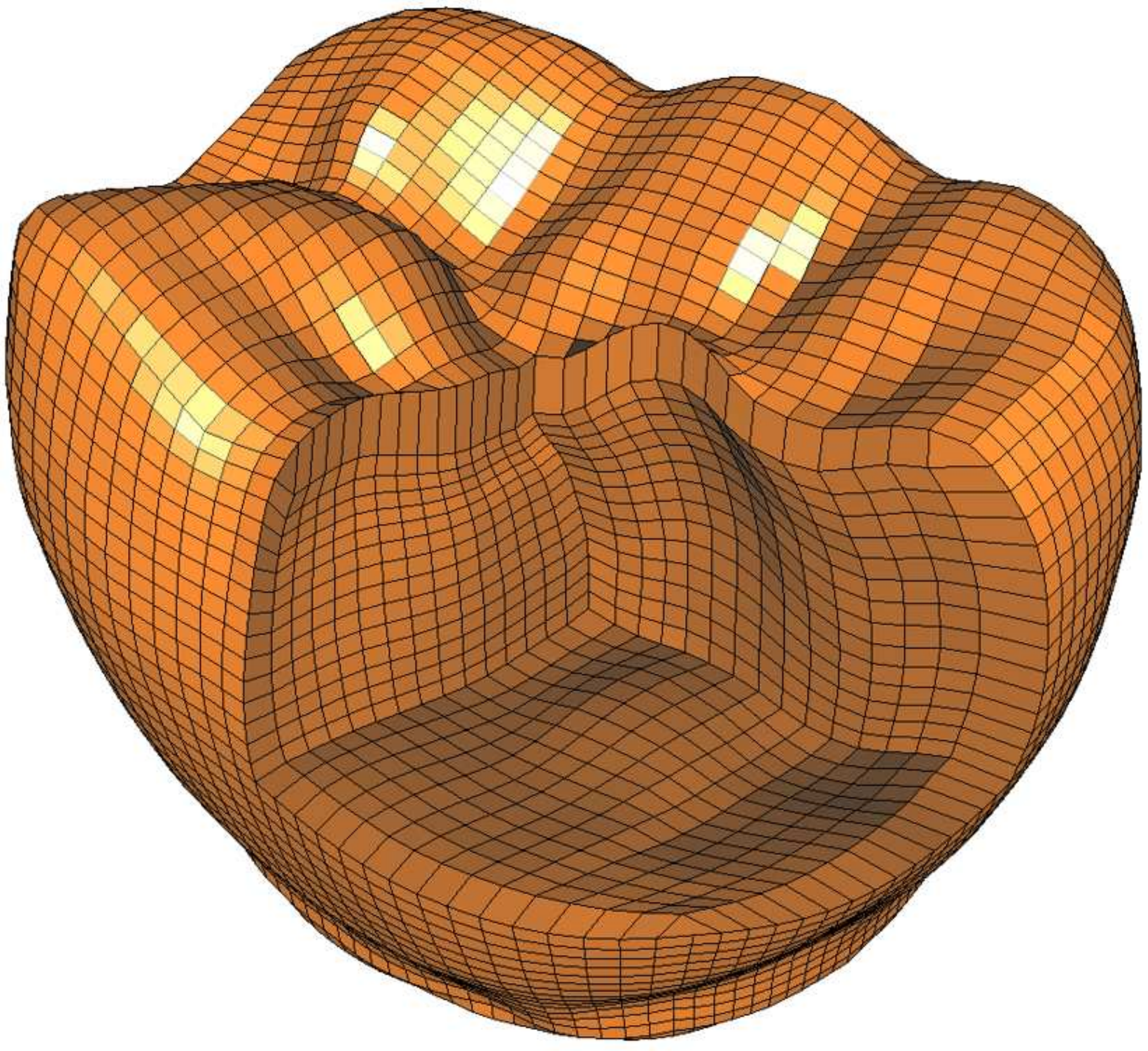}}
  \subfigure[]{
    \label{subfig:headcut}
    \includegraphics[width=0.17\textwidth]{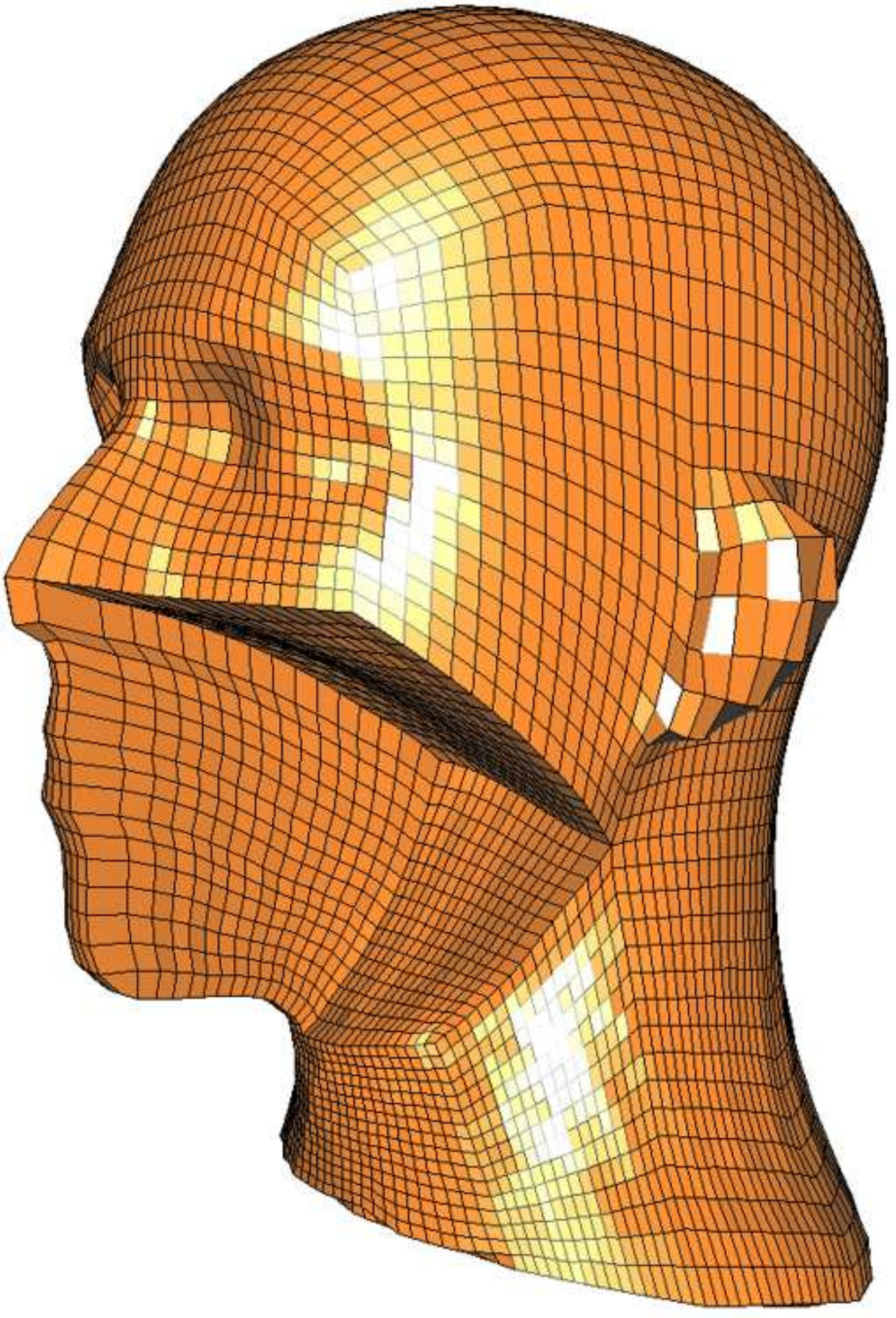}}
  \subfigure[]{
    \label{subfig:moaicut}
    \includegraphics[width=0.14\textwidth]{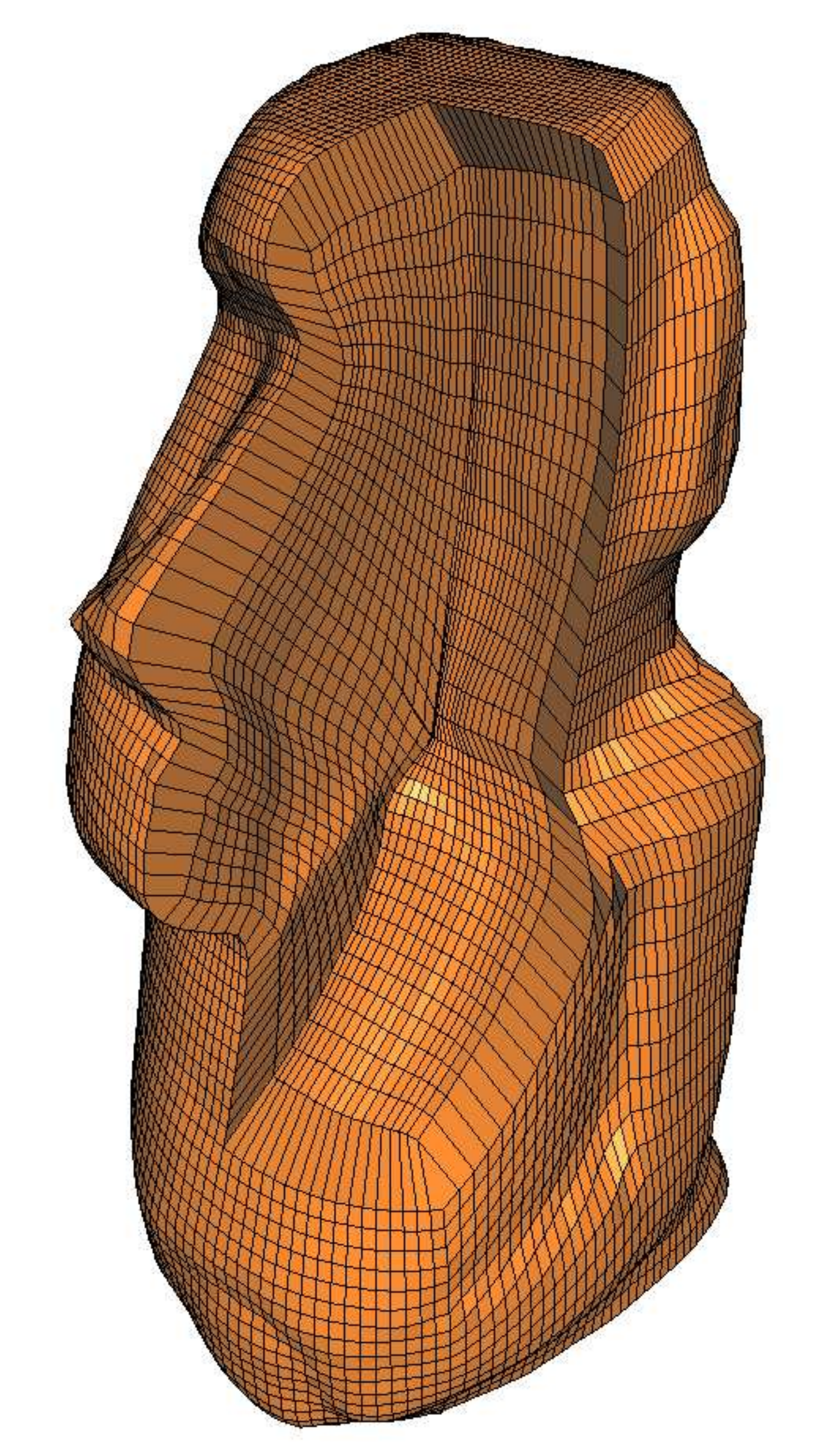}}
      \caption{\small The polyhedral volume parametric mesh in the physical domains \emph{Duck} (a), \emph{Ball joint} (b), \emph{Tooth} (c), \emph{Head} (d), and \emph{Moai} (e), and their cut away views (f), (g), (h), (i), (j).}
   \label{fig:volume_mesh}
  \end{center}
\end{figure}

 In Fig.~\ref{fig:volume_mesh}(a)-(e), the polyhedral volume parametric mesh
    generated by the Gregory solid mapping are illustrated,
    with their cut-away views (Fig.~\ref{fig:volume_mesh}(f)-(j)).
 The average, minimal, and maximal scaled Jacobians of these Gregory solids
    are listed in Table~\ref{tbl:stat}.
 The average scaled Jacobians of the Gregory solids are all above $0.8$.
 Although there are still some regions with negative Jacobians in four
    models,
    the ratio between the volume of region with negative Jacobian and the volume of the whole model are below $0.18\%$ (refer to the column of $J^- / J$ in Table~\ref{tbl:stat}).
 Further checking shows that the regions with negative Jacobian concentrate
    on the regions near the inappropriately segmented boundary patches,
    for example, two adjacent boundary patches are $C^1$ continuous along their common boundary curve.

 Moreover, the last column of Table~\ref{tbl:stat} lists the running time of
    the Gregory solid construction and optimization algorithm,
    ranging from 130.22 seconds to 2386.22 seconds.
 Specially, we can see that the tooth model has no region with negative Jacobian after optimization
    and the running time of optimization is much shorter than others,
    because of the segmentation of tooth model and the initial generated hexahedral model is much better
    than others.
 There is still some space for accelerating the algorithm.
 In Table~\ref{tbl:coef}, the weights $\mu$ and $\nu$ (in the objective
    function~\pref{eq:org_optimization_problem}) employed in generating each Gregory solid are presented.
 As stated above, the weights $\mu$ and $\nu$ are used to balance the values
    of the three items in the objective function~\pref{eq:org_optimization_problem}.
 Because the orders of magnitude of the three items differ in the
    optimization of each Gregory solid,
    the weights $\mu$ and $\nu$ also differ in each optimization.

  \begin{table*}[!htb]
  \centering
  \footnotesize
  \caption{Weights employed in optimization.}
  \label{tbl:coef}
  \begin{threeparttable}
  \begin{tabular}{| c | c | c | c | c |c|c|}
  \hline
              & Duck     &  Ball joint & Tooth    & Head     & Moai\\
  \hline
    $\mu$    & 0.000001  & 0.00001     & 0.00001  & 0.000001 & 0.00002\\
  \hline
   $\nu$     & 0.3       & 0.1         & 0.01     & 0.03    & 0.02  \\
  \hline
  \end{tabular}
 \end{threeparttable}
 \end{table*}


\section{Conclusion}
 \label{sec:conclusion}

 In this paper, we developed the Gregory solid representation,
    and employed it to interpolate the four-sided or non-four-sided boundary patches of a polyhedral physical domain.
 Moreover, the algebraic quality of the Gregory solid is improved by solving
    a sparse optimization problem using the ADMM method.
 In this way, the polyhedral volume parametrization of a given physical
    domain with four-sided or non-four-sided boundary patches can be generated.
 Experiments show that, in the polyhedral volume parametrization produced by
    the Gregory solid construction and sparse optimization method,
    the regions with negative Jacobian are very small (below 0.18\%),
    and they usually concentrate around the boundary curves where two boundary patches are $C^1$ continuously stitched.
 As a future work, we will study how to entirely eliminate the region with
    negative Jacobian by optimizing the boundary patch segmentation.

\section*{Acknowledgement}

 This paper was supported by the National Natural Science Foundation of China (No. 61872316),
   the National Key R\&D Program of China (No. 2016YFB1001501),
   and the Fundamental Research Funds for the Central Universities (No. 2017XZZX009-03).

\bibliographystyle{unsrt}
\bibliography{GregorySolid}

\begin{thebibliography}{10}

\bibitem{Hughes2005Isogeometric}
T.~J.~R. Hughes, J.~A. Cottrell, and Y.~Bazilevs.
\newblock Isogeometric analysis: Cad, finite elements, nurbs, exact geometry
  and mesh refinement.
\newblock {\em Computer Methods in Applied Mechanics \& Engineering},
  194(39):4135--4195, 2005.

\bibitem{Knupp2015Achieving}
Patrick~M Knupp.
\newblock Achieving finite element mesh quality via optimization of the
  jacobian matrix norm and associated quantities. part ii¡ªa framework for
  volume mesh optimization and the condition number of the jacobian matrix.
\newblock {\em International Journal for Numerical Methods in Engineering},
  48(8):1165--1185, 2000.

\bibitem{Boyd2010Distributed}
Stephen Boyd, Neal Parikh, Eric Chu, Borja Peleato, and Jonathan Eckstein.
\newblock Distributed optimization and statistical learning via the alternating
  direction method of multipliers.
\newblock {\em Foundations \& Trends in Machine Learning}, 3(1):1--122, 2010.

\bibitem{Floater1997Parametrization}
Michael~S Floater.
\newblock Parametrization and smooth approximation of surface triangulations.
\newblock {\em Computer Aided Geometric Design}, 14(3):231--250, 1997.

\bibitem{Sander2001Texture}
Pedro~V. Sander, John Snyder, Steven~J. Gortler, and Hugues Hoppe.
\newblock Texture mapping progressive meshes.
\newblock pages 409--416, 2001.

\bibitem{Alliez1970Recent}
Pierre Alliez, Giuliana Ucelli, Craig Gotsman, and Marco Attene.
\newblock Recent advances in remeshing of surfaces.
\newblock {\em Mathematics \& Visualization}, pages 53--82, 1970.

\bibitem{Hormann2000Mips}
K~Hormann.
\newblock Mips : An efficient global parametrization method.
\newblock {\em Curve and Surface Design: Saint-Malo}, 2000.

\bibitem{Heras2003An}
Jos¨¦ Mar¨ªa~Las Heras.
\newblock An adaptable surface parameterization method.
\newblock {\em Proceedings of International Meshing Roundtable}, (7):201--213,
  2003.

\bibitem{Floater2005Surface}
Michael~S. Floater and Hormann Kai.
\newblock {\em Surface Parameterization: a Tutorial and Survey}.
\newblock Springer Berlin Heidelberg, 2005.

\bibitem{Sheffer2006Mesh}
Alla Sheffer, Emil Praun, and Kenneth Rose.
\newblock Mesh parameterization methods and their applications.
\newblock {\em Foundations \& Trends? in Computer Graphics \& Vision},
  2(2):105--171, 2006.

\bibitem{Martin2008Volumetric}
T.~Martin, E.~Cohen, and R.M. Kirby.
\newblock Volumetric parameterization and trivariate b-spline fitting using
  harmonic functions.
\newblock pages 269--280, 2008.

\bibitem{Martin2010Volumetric}
T.~Martin and E.~Cohen.
\newblock Volumetric parameterization of complex objects by respecting multiple
  materials.
\newblock {\em Computers \& Graphics}, 34(3):187--197, 2010.

\bibitem{Lin2015Constructing}
Hongwei Lin, Sinan Jin, Qianqian Hu, and Zhenbao Liu.
\newblock Constructing b-spline solids from tetrahedral meshes for isogeometric
  analysis.
\newblock {\em Computer Aided Geometric Design}, 35-36:109--120, 2015.

\bibitem{Wang2013Trivariate}
Wenyan Wang, Yongjie Zhang, Lei Liu, and Thomas J.~R. Hughes.
\newblock Trivariate solid t-spline construction from boundary triangulations
  with arbitrary genus topology.
\newblock {\em CAD Computer Aided Design}, 45(2):351--360, 2013.

\bibitem{Xu2013Analysis}
Gang Xu, Bernard Mourrain, R~Duvigneau, and Andr Galligo.
\newblock Analysis-suitable volume parameterization of multi-block
  computational domain in isogeometric applications.
\newblock {\em CAD Computer Aided Design}, 45(2):395--404, 2013.

\bibitem{Zhang2013Conformal}
Yongjie Zhang, Wenyan Wang, and Thomas~J. Hughes.
\newblock {\em Conformal solid T-spline construction from boundary T-spline
  representations}.
\newblock Springer-Verlag New York, Inc., 2013.

\bibitem{Wang2014An}
Xilu Wang and Xiaoping Qian.
\newblock An optimization approach for constructing trivariate bb-spline
  solids.
\newblock {\em Computer-Aided Design}, 46(1):179--191, 2014.

\bibitem{Lin2018Trivariate}
Hongwei Lin, Hao Huang, and Chuanfeng Hu.
\newblock Trivariate b-spline solid construction by pillow operation and
  geometric iterative fitting.
\newblock {\em SCIENCE CHINA (Information Sciences)}, 61:232--234, 2018.

\bibitem{Meyer2002Generalized}
Mark Meyer, Alan Barr, Haeyoung Lee, and Mathieu Desbrun.
\newblock Generalized barycentric coordinates on irregular polygons.
\newblock {\em Journal of Graphics Tools}, 7(1):13--22, 2002.

\bibitem{Floater2003Mean}
Michael~S. Floater.
\newblock Mean value coordinates.
\newblock {\em Computer Aided Geometric Design}, 20(1):19--27, 2003.

\bibitem{Floater2005Mean}
Michael~S Floater, G~S, and Martin Reimers.
\newblock Mean value coordinates in 3d.
\newblock {\em Computer Aided Geometric Design}, 22(7):623--631, 2005.

\bibitem{Joshi2007Harmonic}
Pushkar Joshi, Mark Meyer, Tony Derose, Brian Green, and Tom Sanocki.
\newblock Harmonic coordinates for character articulation.
\newblock {\em Acm Transactions on Graphics}, 26(3):71, 2007.

\bibitem{Lipman2008Green}
Yaron Lipman, David Levin, and Daniel Cohen-Or.
\newblock Green coordinates.
\newblock {\em Acm Transactions on Graphics}, 27(3):1--10, 2008.

\bibitem{Chiyokura1983Design}
Hiroaki Chiyokura and Fumihiko Kimura.
\newblock Design of solids with free-form surfaces.
\newblock In {\em Conference on Computer Graphics \& Interactive Techniques},
  pages 289--298, 1983.

\bibitem{Chiyokura1986Localized}
Hiroaki Chiyokura.
\newblock Localized surface interpolation method for irregular meshes.
\newblock In {\em Computer Graphics Tokyo '86 on Advanced Computer Graphics},
  pages 3--19, 1986.

\bibitem{Gregory1974SMOOTH}
John~A. Gregory.
\newblock Smooth interpolation without twist constraints.
\newblock {\em Computer Aided Geometric Design}, pages 71--87, 1974.

\bibitem{Longhi1987Interpolating}
Lucia Longhi.
\newblock Interpolating patches between cubic boundaries.
\newblock {\em Computer Science Division}, 1987.

\bibitem{Wang2004Non}
Charlie C.~L. Wang and Kai Tang.
\newblock Non-self-overlapping structured grid generation on an n-sided
  surface.
\newblock {\em International Journal for Numerical Methods in Fluids},
  46(9):961--982, 2004.

\bibitem{Avriel1976Nonlinear}
M~Avriel.
\newblock {\em Nonlinear programming : analysis and methods}.
\newblock Prentice-Hall, 1976.

\end{thebibliography}

\end{document}